\documentclass[prc,letterpaper,showpacs,preprint]{revtex4}

\usepackage{graphicx}
\usepackage{amsmath}
\usepackage{bm}

\begin{document}

\preprint{ \vbox{\hbox{ JLAB-THY-04-281}}}

\title{A comprehensive treatment of electromagnetic interactions and the
three-body spectator equations}

\author{J.~Adam, Jr.$^{(1)}$, J.~W.~Van Orden$^{(2,3)}$}

\affiliation{\small \sl (1) Nuclear Physics Institute, Czech
Academy of Sciences,
CZ-25068 {\v R}e{\v z} n.\ Prague, Czech Republic \\
 (2) Jefferson Lab,
12000 Jefferson Avenue, Newport News, VA 23606\\
(3) Department of Physics, Old Dominion University, Norfolk, VA
23529 }

\date{\today}

\begin{abstract}
We present a general derivation the three-body spectator (Gross)
equations and the corresponding electromagnetic  currents. As in
previous paper on two-body systems, the wave equations and currents
are derived from those for Bethe-Salpeter equation with the help of
algebraic method using a concise matrix notation. The three-body
interactions and currents introduced by the transition to the
spectator approach are isolated and the matrix elements of the e.m.
current are presented in detail for system of three indistinguishable
particles, namely for elastic scattering and for two and three body
break-up. The general expressions are reduced to the
one-boson-exchange approximation to make contact with  previous work.
The method is general in that it does not rely on introduction of the
electromagnetic interaction with the help of the minimal replacement.
It would therefore work also for other external fields.
\end{abstract}


 \maketitle
%___________________________________________________________________________
\section{Introduction}

Exposing few nucleon systems to medium energy ($\sim 1$GeV)
electromagnetic probes requires a relativistic description of initial
and final states as well as of the reaction mechanism. The relativistic
effects are usually included in perturbative way \cite{vc}, but even at
leading order one has to include large number of terms into NN
potential and electromagnetic current, while the convergence of the
series in $v/c \sim p/m$ becomes questionable. The conceptually cleaner
alternative is to start from an explicitly covariant formulation. A
number of covariant techniques has  been introduced and studied
extensively for the case of the simplest two nucleon system, in
particular in calculations of the elastic e.m.\  deuteron form factors
\cite{GVO} and in deuteron electro- and photodisintegration \cite{GGr}.

In the present paper we adopt the spectator (or Gross)
quasipotential approach \cite{spectator,Gr82}. It reduces four
dimensional integrals over loop momenta to their more treatable
three dimensional form in explicitly covariant way by picking up
the poles of nucleon propagators. This simplifies an analytic
structure of dynamic equations and leads to closer approximation
to full result compared to the Bethe-Salpeter ladder
approximation.  A set of realistic one-boson-exchange potentials
were constructed within the formalism, giving a fair fit to NN
scattering data and deuteron properties \cite{NNspect}. It was
found that the model gives very good description of the elastic
deuteron form factors \cite{GVO,deutff}. Three nucleon spectator
equation with those realistic potentials was  solved by Stadler
and Gross \cite{sg} and the normalization condition for the 3N
bound state amplitudes was derived \cite{norm}. The first
calculation of electromagnetic break-up of 3N bound state are in
progress.

In order to calculate this process it is necessary to determine the
correct effective electromagnetic current operator to be used with
the three-body spectator equation. So far this problem has been
approached in two ways.  The first of these is the elegant method of
Kvinikhidze and Blankleider \cite{kb97,kb97II,kb99} that gauges the
integral equations which describe the strongly interacting system in
order to obtain the electromagnetic current matrix elements for
various processes. The second approach is that of Gross, Stadler and
Pe\~na \cite{Gr04} which uses an approach based on a diagrammatic
description of the three-body system and relies heavily on the
gauging approach. These two approaches lead to identical results once
differences in organization are accounted for. In both of these
papers certain assumptions are inherent in the results that are
presented \cite{Gr04}. In both cases, no explicit three-body forces
are included and the results implicitly assume that the interaction
kernels are given by the one-boson-exchange approximation.

In a previous paper \cite{2Ncur} we pursued a different strategy to
obtain a general derivation of the e.m.\ currents for the two-body
Gross equation. The starting point was the organization of the
corresponding Feynman amplitudes into 2N Bethe-Salpeter equation and
factorization of five point Green function (with the external e.m.\
field). The spectator equations were then obtained by separating the
propagator of one of the two particles into a piece containing only
the on-shell positive-energy contribution and a remainder. The
Bethe-Salpeter amplitudes are then rearranged into integral equations
which contain  only  the on-shell propagator and a quasipotential
equation which sums all of the off-shell contributions. The
electromagnetic current consistent with this rearrangement was
derived with the help  of operator algebra, which defines the
products of singular operators and represents an useful shortcut
alternative to pole analysis of the Feynman diagrams. We have shown
that this definition is consistent with gauge invariance when a
necessary truncation of the strong kernel is applied in {\it any}
finite order of quasipotential decomposition and in a number of meson
exchanges involved. The main aim of the present paper is to extend
this treatment to 3N spectator equation and to derive general
expressions for electromagnetic matrix elements needed for the future
application to actual calculation of the electromagnetic properties
of relativistic three nucleon systems.

For convenience, Sec. II contains a brief review of the application
of this approach to the two-body system.

The starting point of  the  analogous development for three-body
systems is a description of the three-body Bethe-Salpeter amplitudes
given by four-dimensional integral equations that iterate kernels
consisting of all three-body irreducible contributions to the
six-point function. The currents consistent with these kernels can be
obtained either by attaching a photon to all internal particle
propagators in a diagrammatic representation the kernels, or by use
of the gauging method \cite{kb99}. Section III contains a review of
the three-body Bethe-Salpeter equations and current including
contributions from two- and three-body kernels and currents. Since
the three-body spectator equations are defined only for three
identical particles a discussion of the three-body equations for
identical particles is given in Sec.~IIIC. Before using our approach
to obtain the spectator equations, it is convenient to use the
symmetry of the various Faddeev amplitudes to limit the number of
channels that explicitly appear in the equations.  We found that it
is most convenient to do this in the context of a matrix
representation of the channel space. Appendix A describes the matrix
method for three distinguishable particles and Appendix B describes
the case of identical particles and show how to reduce the problem to
two channels.

Section~IV applies the operator technique to the two-dimensional
matrix representation of the Bethe-Salpeter equation to obtain the
corresponding spectator equations for the t-matrix and wave
functions, and obtains an expression for the corresponding effective
current operator. Some care is taken to separate two-body
quasipotential and scattering matrix contributions from three-body
contributions that are generated by the transformation from the
Bethe-Salpeter to the spectator equations. These contributions can
occur even if there are no explicit three-body Bethe-Salpeter
interactions.

In Sec.~V we show how the general expressions obtained in Sec.~IV can
be reduced to the case of the one-boson-exchange approximation for
the Bethe-Salpeter kernels and currents.  This gives results
identical to those of Gross, Stadler and Pe\~na; and, by extension,
to those of Kvinikhidze and Blankleider.

%=============================================================
\section{Review of the Two-body Equations}

The motivation for the arguments that we present below for
constructing the three-body electromagnetic current matrix
elements for the three-body spectator equation is best illustrated
by providing an outline of the arguments that appeared in our
previous paper \cite{2Ncur} for obtaining the equivalent
expressions for the two-body spectator equation.

First consider the two-body Bethe-Salpeter equation for two
particles interacting by the exchange of a boson.  Examination of
all of the diagrams contributing to the two-body t-matrix (the
four-point vertex function) allows to show that the sum of
all contributions to the t-matrix can be represented by an
integral equation of the form
\begin{equation}
 M=V-V G_{BS}  M \, ,
\end{equation}
where the kernel of the equation $V$ is the sum of all two-body
irreducible diagrams, and where $G_{BS}=-iG_1G_2$ is the free
two-body Bethe-Salpeter propagator.
%\footnote{
The formalism reviewed in this section and further developed
in this paper for three particle systems is restricted to the energies below
particle production threshold. That is, the propagation of the constituents is
approximated by free Green's functions with the physical mass and
the phenomenologically dressed interaction vertices are assumed, i.e.,
explicit loops dressing two- and three-point Green's functions and their
renormalization are not considered.
%}
The two-body interacting propagator (four-point propagator) can then be written as
\begin{equation}
{\cal G}=G_{BS}-G_{BS}  M G_{BS}=G_{BS}-G_{BS} V {\cal G}=
G_{BS}-{\cal G} V G_{BS}\,.
\end{equation}
This propagator contains all of the information about the two-body
system and the bound and scattering states wave functions can be
obtained from the singularities of this propagator. By looking for
sub-threshold poles in the propagator the bound state
Bethe-Salpeter wave function can be determined to be
\begin{equation}
\left|\psi\right>=G_{BS}\left|\Gamma\right> \, ,
\end{equation}
where $\left|\Gamma\right>$ is the bound-state Bethe-Salpeter
vertex function which satisfies the equation
\begin{equation}
\left|\Gamma\right>=-VG_{BS}\left|\Gamma\right>\,.
\end{equation}
The pole in the propagator associated with the two incoming
particles being on mass shell gives the scattering wave function
with outgoing wave boundary conditions
\begin{equation}
\left|\psi^{(+)}\right>=\left(1-G_{BS}  M \right)
\left|\bm{p}_1,s_1;\bm{p}_2,s_2\right>
\end{equation}
and the corresponding pole for the outgoing particles gives the
conjugate scattering wave function with incoming wave boundary
conditions
\begin{equation}
\left<\psi^{(-)}\right|=\left<\bm{p}_1,s_1;\bm{p}_2,s_2\right|
\left(1- M G_{BS}\right)\,.
\end{equation}
All of the wave functions satisfy the wave equation
\begin{equation}
{\cal G}^{-1}\left|\psi\right>=0\,.
\end{equation}

The relationship of the two-body interacting propagator to the wave
functions implies that we can obtain all of the necessary information
concerning the electromagnetic currents  for this system in
one-photon-exchange approximation by examining all Feynman diagrams
contributing to a five-point propagator with one photon leg and four
interacting particle legs. This five-point propagator can be written
as
\begin{equation}
{\cal G}^\mu=-\cal{G}J^\mu\cal{G}
\end{equation}
by defining the current operator
\begin{equation}
J^\mu=iJ_1^\mu G^{-1}_2+iJ_2^\mu G^{-1}_1+J^{(2)\mu}_{12} \, ,
\end{equation}
where $J^\mu_i$ is the one-body current operator for particle $i$
and $J^{(2)\mu}_{12}$ is the two-body current composed of all two-body
irreducible diagrams involving the absorption of a photon.  Any
current matrix element can then be obtained by picking out the
poles in the initial and final interacting propagators
corresponding to either bound or scattering states. This current
operator satisfies the Ward-Takahashi identity
\begin{equation}
q_\mu J^\mu=\left[e_1(q)+e_2(q),{\cal G}^{-1}\right]\, ,
\end{equation}
where $q_\mu$ is the photon momentum, $e_t(q)$ is the
product of the charge $e_t$ for particle $t$ (which might be an
operator in isospin space) and a  four-momentum shift operator
defined such that
\begin{equation}
\left<p'_t| e_t(q)|p_t\right>=e_t(2\pi)^4\delta^4(p'_t-p_t-q) \, .
\end{equation}
This along with the wave equation guarantees that the current will
be conserved for all possible two-body matrix elements.

The spectator or Gross equation (for distinguishable particles)
can be obtained from the Bethe-Salpeter
equation by substituting a two-body noninteracting propagator $g_1$
with particle 1 on mass shell for the Bethe-Salpeter propagator
$G_{BS}$. The t-matrix will
remain unchanged if it is written as a pair of coupled equations
\begin{equation}
 M=U-U g_1  M
\end{equation}
and
\begin{equation}
U=V-V \Delta g_1 U \, ,
\end{equation}
where the new kernel $U$ is called the quasipotential and $\Delta
g_1=G_{BS}-g_1$.

In \cite{2Ncur} it was shown that there exist a set of relationships
for the spectator scattering matrix, propagators, wave functions and
currents which are of the same form as those given above for the
Bethe-Salpeter equation and that an effective current operator can be
derived,  satisfying a Ward-Takahashi identity which leads to current
conservation when used with the spectator wave functions. For
identical particles the formalism can be represented in a compact and
transparent matrix notation.

We will now show that a similar organization can be obtained for
the three-body Bethe-Salpeter and Gross equations.  We will do
this assuming that there may be explicit three-body interactions
in addition to two-body interactions and will obtain expressions
for the three-body Gross equations and effective currents which
can be expanded to arbitrary order.

%=============================================================

\section{Three-body Bethe-Salpeter Equation}

%______________________________________________________________
\subsection{Three-body t-matrix and wave functions}

The three-body Bethe-Salpeter equation can be motivated in much the
same way as in the two-body case. Examining the Feynman diagrams
contributing to the three-body t-matrix it is possible to separate them
into two categories, three-body reducible and irreducible diagrams.
The irreducible diagrams fall into two classes, those where only two of
the three particles are interacting and those where all three particles
are interacting. The sum of all irreducible diagrams can then be
written as
\begin{equation}
 {\cal V} = V^0 + \sum_{i= 1}^3 V^i iG^{-1}_i =
 {\cal V}^0 + \sum_{i= 1}^3 {\cal V}^i \, ,
 \label{Vkernel}
\end{equation}
where $V^0$ is the sum of all three-body irreducible diagrams where
all three particles are interacting and $V^i$ is the sum of all
diagrams where only particles $j$ and $k$ (with $j\neq k\neq i$) are
interacting. That is $V^0$ is a three-body interaction and $V^i$ is a
two-body interaction. The three-body t-matrix can then be written as
\begin{equation}
{\cal T}= {\cal V}- {\cal V}G^0_{BS}{\cal T}=
          {\cal V}- {\cal T}G^0_{BS}{\cal V} \, ,
\end{equation}
where the three-particle free propagator in terms of the individual
propagators reads $G^0_{BS}=-G_1G_2G_3$.

The three-body Bethe-Salpeter six-point Green function is then
given as:
\begin{eqnarray}
{\cal G}&=& G^0_{BS}-G^0_{BS}{\cal V}{\cal G}=
            G^0_{BS}-{\cal G}{\cal V}G^0_{BS} \nonumber\\
        &=& G^0_{BS}-G^0_{BS}{\cal T}G^0_{BS} \, .
\label{Gresolved}
\end{eqnarray}
Obviously, the inverse three-body propagator is
\begin{equation}
{\cal G}^{-1}=(G^0_{BS})^{-1}+{\cal V} \, .
\end{equation}

It is convenient to introduce the Faddeev decomposition of the full t-matrix
into subamplitudes ${\cal T}^{ij}$,
\begin{equation}
{\cal T}= \sum_{i,j=0}^3 {\cal T}^{ij} \, ,
\label{Tfad}
\end{equation}
where the indices $i$ and $j$ characterize the subamplitudes according
to the character of the last and first interactions; ie., for $i\neq 0$
($j\neq 0$) , the particles $i$ ($j$) are not taking part in the last
(first) interaction, $i=0$ ($j=0$) means that the last (first)
interaction is genuine  three-particle interaction.
The subamplitudes ${\cal T}^{ij}$ satisfy the following integral
equations (see also ref.\ \cite{norm})
\begin{equation}
{\cal T}^{ij}= {\cal V}^i\delta_{ij}- {\cal V}^i G^0_{BS}
 \sum_{k=0}^3  {\cal T}^{kj}\,.
\end{equation}
For numerical solution it is more appropriate to re-write these
equations in terms of two-body and three-body t-matrices which
subsume the corresponding interaction kernels \cite{sg}:
\begin{equation}
 M^i=V^i-V^iG^i_{BS} M^i \, ,
\label{m-two}
\end{equation}
where $M^i$ is the two-body scattering matrix and  $G^i_{BS}=-iG_jG_k=
iG^{-1}_iG^0_{BS}$;  and
\begin{equation}
 M^0=V^0-V^0G^0_{BS} M^0 \, ,
\label{m-three}
\end{equation}
iterates to all orders the three-body interaction . The subamplitudes
for three-body Bethe-Salpeter equation for distinguishable particles
can be written as
\begin{eqnarray}
{\cal T}^{ij}&=&  M^i iG^{-1}_i\delta_{ij}-  M^i G^i_{BS} \sum_{k=0}^3
B_{ik} {\cal T}^{kj} \nonumber\\
&=&{\cal M}^i \delta_{ij}- {\cal M}^i G^0_{BS} \sum_{k=0}^3 B_{ik}
{\cal T}^{kj} \, ,
\label{three-body_BS}
\end{eqnarray}
where $B_{ik}=1-\delta_{ik}$ and ${\cal M}^i$ are defined in terms of
$M^i$ as in eq.\ (\ref{Vkernel}).

As in the two-body case, the three-body Bethe-Salpeter wave functions
can be obtained by examining the residues of poles in the interacting
three-body propagator. The wave function for a three-body bound state
with mass $M_B$ is related to the residue of the propagator at a pole
corresponding to $P^2= M_B^2$, where $P$ is the total momentum of the
three-body system. From the resolved form of the propagator, it is
clear that this is due to a pole in the t-matrix. The three-body
bound state vertex function is related to the residue of the t-matrix
at this pole so that the t-matrix reads
\begin{equation}
{\cal T}= - \frac{\left|\Gamma\right>
\left<\Gamma\right|}{M_2^2-P^2}+{\cal R} \, .
\end{equation}
Using this in the t-matrix equation, the vertex function can be
shown to satisfy the equation
\begin{equation}
 \left|\Gamma\right>= - {\cal V}  G^0_{BS} \left|\Gamma\right> \, .
\end{equation}
The interacting three-body propagator can be written as
\begin{equation}
{\cal G}=\frac{G^0_{BS}\left|\Gamma\right>
\left<\Gamma\right|G^0_{BS}}{M_2^2-P^2}+G^0_{BS}-G^0_{BS}{\cal
R}G^0_{BS} \, .
\end{equation}
The bound-state wave function is defined as
\begin{equation}
|\Psi >= G^0_{BS} \left|\Gamma\right> \, ,
\end{equation}
which satisfies the wave equation
\begin{equation}
 {\cal G}^{-1}|\Psi >=0 \, .
\end{equation}
The bound state vertex and wave functions are decomposed into their
Faddeev components:
\begin{eqnarray}
 |\Gamma >&=& \sum_{i=0}^3 |\Gamma^i > \, ,\\
 |\Psi >&=& \sum_{i=0}^3 |\Psi^i > \, ,
\end{eqnarray}
defined by dynamical equations
\begin{eqnarray}
 |\Gamma^i  >&=& - {\cal V}^i G^0_{BS} |\Gamma > \, , \\
 (G^0_{BS})^{-1}\, |\Psi^i >&=&  - {\cal V}^i |\Psi > \, ,
\end{eqnarray}
where $i= 0, 1, 2, 3$.

The scattering states can be obtained from (\ref{Gresolved}).
Since $G^0_{BS}$ has poles at the on-shell masses of particles
with plane-wave residues, isolating the poles for the right-hand
propagators in (\ref{Gresolved}) gives the three-body scattering
wave function with outgoing-wave boundary conditions
\begin{equation}
\left|\Psi^{(+)}\right>=\left(1-G^0_{BS}{\cal T}
\right)\left|\bm{p}_1,s_1;\bm{p}_2,s_2;\bm{p}_3,s_3\right>\,,
\label{scat3BSin}
\end{equation}
while doing so for the left-hand yields the conjugate scattering
wave function with incoming-wave boundary conditions.
\begin{equation}
\left<\Psi^{(-)}\right|=\left<\bm{p}_1,s_1;\bm{p}_2,s_2;\bm{p}_3,s_3\right|
\left(1-{\cal T} G^0_{BS}\right)\,, \label{scat3BS}
\end{equation}
where
\begin{equation}
\left|\bm{p},s\right>=\left\{
\begin{array}{ll}
u(\bm{p},s)\left|\bm{p}\right> & \quad{\rm for\ spin}=\frac{1}{2} \\
\left|\bm{p}\right> &\quad {\rm for\ spin}=0
\end{array}
\right.
\end{equation}
The Faddeev
components of the three-body scattering wave function are given by
\begin{equation}
\left<\Psi^{(-)i}\right|=\left<\bm{p}_1\bm{p}_2\bm{p}_3\right|Q_1Q_2Q_3
\left(\frac{1}{4}- \sum_{k}{\cal T}^{ki} G^0_{BS}\right) \, .
\label{scat3BSf}
\end{equation}

In order to extract the scattering states where two of the particles
are bound and the other is in the continuum, the propagator eq.\ (\ref{Gresolved})
is rewritten using (\ref{three-body_BS}) as
\begin{eqnarray}
{\cal G}&=& G^0_{BS}-G^0_{BS}\sum_{j=0}^3 {\cal T}^{0j}G^0_{BS}
\nonumber\\
&&-\sum_{i=1}^3\sum_{j=0}^3\left[G^i_{BS}{\cal M}^i
iG_iG^i_{BS}\delta_{ij}-G^i_{BS}{\cal M}^iiG_i G^i_{BS}
\sum_{k=0}^3 B_{ik} {\cal T}^{kj}G^0_{BS}\right]\, .
\end{eqnarray}
The last term in this expression contains the two-body scattering
matrices which have poles corresponding to the two-body bound states
and can be represented by
\begin{equation}
{\cal M}^i= - \frac{\left|\Gamma^{(2)i}\right>
\left<\Gamma^{(2)i}\right|}{M_2^2-{P^i}^2}+{\cal R}
\end{equation}
where $\left|\Gamma^{(2)i}\right>$ is the two-body bound-state vertex
function for the pair of interacting particles $j$ and $k$ ($j\neq
k\neq i$), $M_2$ is the mass of the two-body bound state and $P^i$ is
its total momentum. This term also contains the free propagator for
particle $i$. By selecting the residues of the pole for particle 1
(putting $i=1$)  and the bound-state of particles 2 and 3, the
conjugate state two-body scattering state with incoming-wave boundary
conditions can be identified as
\begin{eqnarray}
\left<\Phi^{1(-)}\right|&=&\left<\Gamma^{(2)1};\bm{p}_1,s_1\right|
G^1_{BS}\left[1-\sum_{k=0}^3 \sum_{i=0}^3 B_{1k}
{\cal T}^{ki}G^0_{BS}\right]
\nonumber\\
&=&\left<\Phi^{(2)1};\bm{p}_1,s_1\right|\left[1-\sum_{k=0}^3
\sum_{i=0}^3 B_{1k} {\cal T}^{ki}G^0_{BS}\right]\, ,
\label{scat2BS}\\
\left<\Phi^{1(-)}\right|&=& \sum_{i=0}^3 \left<\Phi^{1(-) i}\right|\, , \\
\left<\Phi^{1(-) i}\right|&=&
\left<\Phi^{(2)1};\bm{p}_1,s_1\right|\left[ \delta_{1i} - \sum_{k
\neq 1} {\cal T}^{ki}G^0_{BS}\right]\, .
\end{eqnarray}
A similar expression can be obtained for the wave function with
outgoing-wave boundary conditions.

All of the wave functions satisfy the wave equation
\begin{equation}
{\cal G}^{-1}|\Psi>=0 \,.
\label{BSwf}
\end{equation}

%______________________________________________________________
\subsection{Electromagnetic current in Bethe-Salpeter framework}

The three-body Bethe-Salpeter current can be determined by
considering all diagrams contributing to the seven-point function
with six legs corresponding to the three incoming and outgoing
particles and one photon leg. By separating the diagrams into
three-particle reducible and irreducible contributions the current
operator can be identified as the sum of all irreducible
seven-point functions. There will be three types of contributions:
one-body contributions where the photon attaches to only one of
the interacting particles, two-body contributions where the photon
attaches internally to a two-body interaction, and three-body
contributions where the photon attaches internally to a three-body
interaction.

The one-body currents are of the form
\begin{equation}
J^{t\mu}= -J^\mu_t G_r^{-1}G_s^{-1} \, ,
\label{IAcurr}
\end{equation}
where $t=1,2,3$, $r\neq s\neq t$ and $J^\mu_t$ is the vertex for
attaching a photon to particle $t$, for which $q_\mu
J^\mu_t=\left[e_t(q), G_t^{-1}\, \right]$. Each of these currents
satisfies the Ward-Takahashi identities
\begin{equation}
q_\mu J^{t\mu}=\left[e_t(q), (G^0_{BS})^{-1}\, \right] \, .
\end{equation}

As in the case of two-particle system \cite{2Ncur,GrossRiska}, the
two- and three-body currents follow from attaching a photon line
to all particle lines and into momentum-dependent vertices
internal to the two- and three-body Bethe-Salpeter kernels.
Defining $J^{(2)\mu}_{rs}$ as the two-body current as extracted from
the two-body problem and $J^{(3)\mu}$ as the completely connected
three-body current, the exchange current for the three-body
system can be written as
\begin{equation}
J^{t\mu}_{\rm ex}=
\left\{ \begin{array}{ll}
J^{(2)\mu}_{rs} iG^{-1}_t, &t=1,2,3 \ \ (r\neq s\neq t) \\
J^{(3)\mu},           &t=0
\end{array}\right.
\end{equation}
which satisfies the Ward-Takahashi identities
\begin{equation}
q_\mu J^{t\mu}_{\rm ex}=\left\{ \begin{array}{ll}
\left[e_r(q)+ e_s(q), {\cal V}^t\right], &t=1,2,3\ \ (r\neq s\neq t)\\
\left[e_T(q), {\cal V}^0\right],         &t=0
\end{array}\right.
\end{equation}
where $e_T(q)= e_1(q)+ e_2(q)+ e_3(q)$.

\begin{figure}
\centerline{\includegraphics[height=2in]{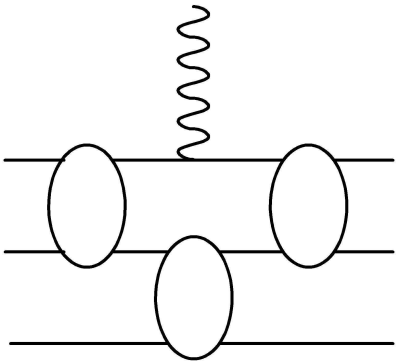}}
\caption{This
Feynman diagram represents part a contribution to the seven-point
function. The particles are label 1 to 3 from top to bottom. The
ovals represent two-body scattering matrices.}
\label{FigA}
\end{figure}

Summation of the diagrams to give the seven-point function must be
done with considerable care \cite{kb97}. The three-body
Bethe-Salpeter equation is based on Feynman perturbation theory
rather than on time-ordered perturbation theory as in the
non-relativistic case. So far this difference has been manifest
only in the need to introduce inverse propagators into some of the
operators in order to obtain equations with a similar structure to
the non-relativistic case and in the fact that the implied
integrals are four-dimensional rather than three-dimensional. The
problem with seven-point function can be seen by considering
Fig.~\ref{FigA}.
%
% Fig 1
Here the virtual photon is absorbed on particle 1 and is preceded by
two-body interactions between particles 1 and 2. Between these
interactions is two-body t-matrix for particles 2 and 3. Since this
is a Feynman diagram, there is no ordering implied between the
absorbed photon and the t-matrix for particles 2 and 3. In summing
the interactions this t-matrix can either be associated with the sum
on the left or with the sum on the right, but not with both.  As a
result, if we associate this interaction with the right hand sum the
last interaction to the right must be between particles two and three
while the first interaction on the left must be between particles 1
and 2. This requires that there must be a restricted sum between
three-body scattering matrices which occur on either side of a
one-body current. The possibility of double-counting this
contribution was pointed out in \cite{kb97II}. The seven-point
function can then be written as
\begin{eqnarray}
{\cal G}^\mu&=&-\sum_{t=1}^3 \left(
G^0_{BS}-G^0_{BS}\sum_{i,j=0}^3{\cal T}^{ij}
B_{jt}G^0_{BS}\right)J^{t\mu} \left(
G^0_{BS}-G^0_{BS}\sum_{k,l=0}^3{\cal T}^{kl}
G^0_{BS}\right)\nonumber \\
& &-{\cal G}\left(\sum_{t=0}^3 J^{t\mu}_{\rm ex} \right){\cal G} \, ,
\label{seven-point_BS1}
\end{eqnarray}
where the restricted sum is represented by the $B_{jt}$ associated with
the left-hand t-matrix. While the exchange current term is nicely
factored with complete interacting six-point propagators on the left
and right of the current operator, the one-body current term is not
factored in this way. This factorization can be accomplished by using
the identity
\begin{equation}
{\cal G}^{-1}\left( G^0_{BS}-G^0_{BS}\sum_{i,j=0}^3{\cal T}^{ij}
B_{jt}G^0_{BS}\right)=1+{\cal V}^t G^0_{BS} = 1+ V^t iG^t_{BS} \,,
\end{equation}
to rewrite the seven-point propagator (\ref{seven-point_BS1}) as
\begin{eqnarray}
{\cal G}^\mu&=& - {\cal G}
\left( \sum_{t=1}^3 \left(
  ( 1+{\cal V}^t G^0_{BS})J^{t\mu}+
    J^{(2)\mu}_t G_t^{-1}
   \right)
 + J^{(3)\mu} \right) {\cal G} \nonumber\\
&=& - {\cal G} \left[ \sum_{t=1}^3 J^{t\mu}+
\sum_{t=0}^3 J_{\rm int}^{t\mu} \right]
{\cal G} \, ,
\end{eqnarray}
where we have defined the  one-body and interaction currents in
three-particle space $J^{t\mu}$ and $J_{\rm int}^{t\mu}$, respectively,
by eq.\ (\ref{IAcurr}) and by
\begin{eqnarray}
J_{\rm int}^{t\mu}&=& J^{t\mu}_{ex}+ iV^t J^\mu_t \, \quad
\mbox{for}\ t\neq0 \, ,
\label{Jtint}\\
J_{\rm int}^{0\mu}&=&J_{\rm ex}^{0\mu}= J^{(3)\mu} \, ,
\label{J0int}
\end{eqnarray}
since
\begin{equation}
{\cal V}^t G^0_{BS}J^{t\mu}= iV^t J^\mu_t
\end{equation}
behaves like a two-body current.
%
% Fig 2
\begin{figure}
\centerline{\includegraphics[height=2in]{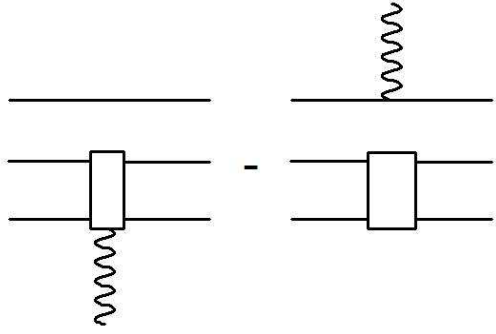}}
\caption{A diagrammatic representation of the interaction current resulting
from the symmetric factorization of the seven-point function. The
rectangular box represents the two-body Bethe-Salpeter kernel and
the rectangular box with attached photon represents the
corresponding two-body electromagnetic current.}
\label{FigB}
\end{figure}
Thus, the total Bethe-Salpeter current is identified as
\begin{equation}
J^\mu_{\rm eff}=\sum_{t=1}^3
J^{t\mu}+\sum_{t=0}^3 J^{t\mu}_{\rm int} = J^{(1)\mu}+ J^\mu_{\rm int} \, ,
\end{equation}
and the WT identity for this current can be easily shown to be
 \begin{equation}
q_\mu J^\mu_{\rm eff} = \left[e_T(q),{\cal G}^{-1}\right] \, .
\label{WT}
\end{equation}
This along with the wave equation for the bound and scattering
states implies that matrix elements of this current between any
three-body states is gauge invariant:
\begin{equation}
 q_\mu {\cal T}^\mu = q_\mu <\Psi'|\, J^\mu_{\rm eff}\, |\Psi>= 0  \, .
\label{emampl}
\end{equation}

%______________________________________________________________
\subsection{Identical particles}

For the case of identical particles it is necessary to explicitly
symmetrize the n-point functions. In particular, the transition to
the corresponding functions for the Gross equation is simplified by
writing the n-point functions in terms of explicitly symmetrized
interactions. Any full operator (summed over all three particles)
${\cal R}$ (e.g. ${\cal R}=  {\cal V}, {\cal T}, {\cal G},
J^{(1)\mu}, J^\mu_{\rm int}, J^\mu_{\rm eff},\dots$) commutes with
any particle interchange operator ${\cal P}_{ij}$: ${\cal R}= {\cal
P}_{ij} {\cal R} {\cal P}_{ij}$. Therefore we can write for its
symmetrized form $R$
\begin{eqnarray}
 R&=& {\cal A}_3 {\cal R}= {\cal R}{\cal A}_3=
 {\cal A}_3 {\cal R}{\cal A}_3 \, , \\
 R&=& {\cal A}_3 R= R {\cal A}_3 \, ,
\end{eqnarray}
where the normalized three-body symmetrization operator is
\begin{eqnarray}
{\cal A}_3\equiv&& \frac{1}{3!}\left( 1+\zeta{\cal P}_{12}+
\zeta{\cal P}_{13}+ \zeta{\cal P}_{23}+
{\cal P}_{4}+{\cal P}_{5}\right)\nonumber\\
=&& \frac{1}{3!}\left( 1+\zeta{\cal P}_{ij}\right)
\left(1+\zeta{\cal P}_{ik}+ \zeta{\cal P}_{jk}\right)=
{\cal A}^k_2\, \Pi_k \nonumber\\
=&& \frac{1}{3!}\left(1+\zeta{\cal P}_{ik}+ \zeta{\cal P}_{jk}\right)
\left( 1+\zeta{\cal P}_{ij}\right)
= \Pi_k\, {\cal A}^k_2 \, ,
\label{symmetry3}
\end{eqnarray}
with $\zeta=1$ ($\zeta=-1$) for bosons (fermions), respectively, and
with
\begin{eqnarray}
{\cal P}_{4}=&&{\cal P}_{13}{\cal P}_{12}=
{\cal P}_{23}{\cal P}_{13}
={\cal P}_{12}{\cal P}_{23} \, , \nonumber\\
{\cal P}_{5}=&&{\cal P}_{23}{\cal P}_{12}={\cal P}_{12}{\cal
P}_{13} ={\cal P}_{13}{\cal P}_{23} \, ,
\end{eqnarray}
and with ${\cal A}^k_2$ being the two body symmetrization operator
acting in the two-body subspaces of particles $i$ and $j$ (where
$i \neq j \neq k$) and the sum of additional exchanges mixing this
subspace with other channel, which we denote $\Pi_k$:
\begin{eqnarray}
{\cal A}^k_2 &=& \frac{1}{2} \left( 1+\zeta{\cal P}_{ij}\right) \, ,
\label{A2k}\\
\Pi_k &=& \frac{1}{3}\left(1+\zeta{\cal P}_{ki}+ \zeta{\cal P}_{kj}\right)
\, .
\label{Pik}
\end{eqnarray}
Note that ${\cal A}_3={\cal A}_3{\cal A}^k_2={\cal A}^k_2{\cal A}_3$
for {\it any\/} $k$. We will exploit this property in the following
discussion.

The symmetrized t-matrix reads
\begin{equation}
T={\cal A}_3{\cal T}= V- V G^0_{BS} T=
 V -T G^0_{BS}  V \, ,
\label{tmatsym}
\end{equation}
with $V$ being the symmetrized potential:
\begin{equation}
 V = {\cal A}_3 {\cal V} = {\cal A}_3 V^0+
\sum_{i=1}^3 \Pi_i\, \overline{V^i} i G^{-1}_i \, ,
\label{Vsym}
\end{equation}
where $\overline{V^i}= {\cal A}^i_2 V^i$ is the symmetrized two particle
interaction. The symmetrized six-point function can then be written as
\begin{equation}
G={\cal A}_3{\cal G}={\cal A}_3 G^0_{BS}-G^0_{BS} T G^0_{BS}
={\cal A}_3 G^0_{BS}-G^0_{BS}  V G
={\cal A}_3 G^0_{BS}- G V G^0_{BS} \, .
\label{Gsym}
\end{equation}
From eqs.\ (\ref{tmatsym},\ref{Gsym}) the symmetrized bound and
scattering state wave functions follow as in previous section. It is
easy to see that for full wave functions it holds
\begin{equation}
 |\Psi_s>= {\cal A}_3 |\Psi> \, .
\end{equation}
One can also define the symmetrized Faddeev components of wave
functions:
\begin{eqnarray}
 |\Psi_s^0>&=& {\cal A}_3 |\Psi^0> \, , \\
 |\Psi_s^i>&=& \frac{1}{3} \left( \, {\cal A}^i_2 |\Psi^i>+
 \zeta{\cal P}_{ij} {\cal A}^j_2 |\Psi^j> +
 \zeta{\cal P}_{ik} {\cal A}^k_2 |\Psi^k >\, \right) \, ,
\end{eqnarray}
where $i \neq j \neq k$.
Making use of the symmetry relations (\ref{Syms}) the symmetrized wave
functions is expressed in terms of only two of these components:
\begin{equation}
|\Psi_s>= |\Psi_s^0>+
(1+ \zeta {\cal P}_{12}+ \zeta {\cal P}_{13}) |\Psi_s^1>=
|\Psi_s^0>+ 3\Pi_1 |\Psi_s^1> \, .
\label{wfsym}
\end{equation}
The symmetrized seven-point function can be written as
\begin{equation}
G^\mu={\cal A}_3{\cal G}^\mu= -G\, J^\mu_{\rm eff}\, G \, ,
\end{equation}
and the symmetrized current matrix elements are
\begin{eqnarray}
 T^\mu = <\Psi_s'|\, J^\mu_{\rm eff}\, |\Psi_s> &=&
 <\Psi_s^{'0}|\, J^\mu_{\rm eff}\, |\Psi_s^0>+
 3 <\Psi_s^{'0}|\, J^\mu_{\rm eff}\, |\Psi_s^1>+ \nonumber\\
 && 3 <\Psi_s^{'1}|\, J^\mu_{\rm eff}\, |\Psi_s^0>+
 3 <\Psi_s^{'1}|\, J^\mu_{\rm eff}\,(1+ 2\zeta {\cal P}_{12}) |\Psi_s^1> \, ,
\label{emamplsym}
\end{eqnarray}
where we have used the symmetry relations for the Faddeev
components of wave functions and the fact that the current
commutes with any replacement operator ${\cal P}_{ij}$.

The BS formalism can be conveniently represented in the matrix notation
with matrix indices corresponding to the Faddeev components. This
matrix notation is particularly useful for further re-arrangement
leading to the spectator formalism. We give a detail description of
this matrix notation in appendices.

%___________________________________________________________________________

\section{The Three-Body Gross Equation}

\subsection{T-matrix and wave functions}

The three-body Gross equations can be obtained from the three-body
Bethe-Salpeter equations by placing two of the intermediate-state
particles on their respective positive energy mass shells. For
amplitudes whose first or last interactions are two-body
interactions, one of these on-shell particles must be the
spectator\cite{Gr82}. For amplitudes whose first of last
interaction are three-body interactions any pair of on-shell
particle can be selected\cite{Gr82}.  This prescription will work
only for three identical particles since it precludes the
possibility that the on-shell pair can form a bound state. Since
the choice of two of the particles to be set on shell clearly
violates the quantum symmetry of the three-body system, it is
necessary to symmetrize the interaction kernels to impose the
appropriate symmetry of the system, as was done in the case of the
two-body system \cite{2Ncur}.

The first step in obtaining the three-body Gross equation is to
obtain the three-body Bethe-Salpeter equation
with symmetrized interactions.  This can be done conveniently by
first expressing the Faddeev components of the various three-body
Bethe-Salpeter amplitudes in terms of four-dimensional matrices
and vectors as described in Appendix A. The symmetrization can
then also be described in terms of matrices as described in
Appendix B. Once the symmetrization is accomplished, it can be
noted that the various Faddeev amplitudes can all be expressed in
terms of only two types of amplitudes allowing the various
amplitudes to be written in terms of two-by-two matrices where the
initial (final) interaction is either a three-body interaction, or
an interaction between particles 2 and 3. This is the 2-channel
formalism of Appendix B. The three-body Gross
equation can now be obtained from this two-dimensional matrix form
by constraining particles 1 and 2 to the positive-energy mass
shell. It can be done conveniently using the operator formalism
introduced in \cite{2Ncur}.

As described in \cite{2Ncur}, a particle can be placed on shell by
picking up only the pole of the propagator for this particle when
performing energy-loop integrals for the Bethe-Salpeter equation.
This can be done by carefully examining these integrals for the
various contributions to a given amplitude or by examining all of
the Feynman diagrams associated with each amplitude. This can be
rather tedious. In \cite{2Ncur}, we developed an algebraic
approach where various products of operators are defined such that
they reproduce the results obtained by choosing poles in the
energy integrals. This approach has the additional advantage that,
by starting from the corresponding Bethe-Salpeter amplitudes it is
possible to obtain expressions for the Gross equation amplitudes
that can be easily approximated to any order in various expansion
schemes. This was shown in detail for the two-body case in
\cite{2Ncur}. The primary step in this approach is to replace the
propagators for the on-shell nucleons using the identity
\begin{equation}
G_i=i{\cal Q}_i+\Delta G_i \, ,
\end{equation}
where ${\cal Q}_i$ is the operator that places the particle on
shell and $\Delta G_i$ is the difference between the full
propagator and its on-shell contribution. Note that $\Delta G_i$
is simply the one-body Feynman propagator with the wrong boundary
condition.

Starting with the two-dimensional forms of the Bethe-Salpeter
amplitudes, we replace the three-body Feynman propagator with the
sum of a piece with particles 1 and 2 on shell $g^1_2{\cal Q}_1
{\cal Q}_2$ and a remainder $\Delta g^1_2$ using
\begin{eqnarray}
G^0_{BS}&=&-G_1 G_2 G_3=-(i {\cal Q}_1+\Delta G_1)
(i {\cal Q}_2+\Delta G_2) G_3
\nonumber\\
&=&g^1_2{\cal Q}_1 {\cal Q}_2+\Delta g^1_2 \, ,
\label{grsub} \\
g^1_2&=& G_3 \, , \\
\Delta g^1_2&=&\left( -i {\cal Q}_1\Delta G_2 -\Delta G_1
i {\cal Q}_2 -\Delta G_1\Delta G_2\right)  G_3 \, .
\end{eqnarray}
Substituting this into the two-dimensional form of the
Bethe-Salpeter scattering matrix (\ref{tmatv2}) gives
\begin{equation}
\bm{t}= \bm{v}-  \bm{v}\, \bm{\delta}\, (g^1_2{\cal Q}_1 {\cal
Q}_2 +\Delta g^1_2)\, \bm{t} \, ,
\end{equation}
where the matrix $\bm{\delta}$ is defined by (\ref{delta}). This
can in turn be reexpressed as a pair of coupled equations as
\begin{eqnarray}
\bm{t}&=&\bm{u}-\bm{u}g^1_2{\cal Q}_1 {\cal Q}_2\bm{\delta}\,
\bm{t}=\bm{u}-\bm{t}g^1_2{\cal Q}_1 {\cal
Q}_2\bm{\delta}\,\bm{u} \,,\label{toff} \\
\bm{u}&=&  \bm{v}-  \bm{v}\, \Delta g^1_2 \bm{\delta}\, \bm{u} =
\bm{v}-  \bm{u}\, \Delta g^1_2 \bm{\delta}\, \bm{v}\, ,
\label{uoff}
\end{eqnarray}
where $\bm{u}$ is the quasipotential.

This system of equations is closed by multiplying from left and
right by ${\cal Q}_1 {\cal Q}_2$ which leads to
\begin{equation}
\bm{t}^1_2=\bm{u}^1_2-\bm{u}^1_2\, g^1_2\, \bm{\delta}\,
\bm{t}^1_2=\bm{u}^1_2-\bm{t}^1_2\, g^1_2\, \bm{\delta}\,
\bm{u}^1_2\, , \label{tgr}
\end{equation}
where $\bm{u}^1_2={\cal Q}_1 {\cal Q}_2\bm{u}{\cal Q}_1 {\cal
Q}_2$ and $\bm{t}^1_2={\cal Q}_1 {\cal Q}_2\bm{t}{\cal Q}_1 {\cal
Q}_2$.

Although (\ref{tgr}) appears to be much like the three-body
Beth-Salpeter equation (\ref{tmatv2}), it is in fact much more
complicated due to the quasipotential equation (\ref{uoff}). The
appearance of the matrix $\bm{\delta}$ in (\ref{uoff}) means that,
in general, the quasipotential matrix is also nondiagonal whereas
the Bethe-Salpeter kernel $\bm{v}$ is diagonal. In accordance with
(\ref{ssym2}) we denote
\begin{equation}
\bm{u}=\left(
\begin{array}{cc}
U^{00} & U^{01}\\ U^{10} & U^{11}
\end{array}
\right)\, .
\label{usym2}
\end{equation}
The off-diagonal
components of $\bm{u}$ are the result of mixing contributions from
the two- and three-body Bethe-Salpeter kernels which results in
contributions which are also completely connected and behave as
three-body interactions. The diagonal elements will also be
complicated by $\bm{\delta}$. The upper left-hand component
$U^{00}$ will always contain at least one contribution from the
three-body Bethe-Salpeter kernel in all of its terms, so it will
always  behave as a three-body interaction even
though contributions from the two-body Bethe-Salpeter kernel will
also occur.

%
% Fig 3
\begin{figure}
\centerline{\includegraphics[height=3in]{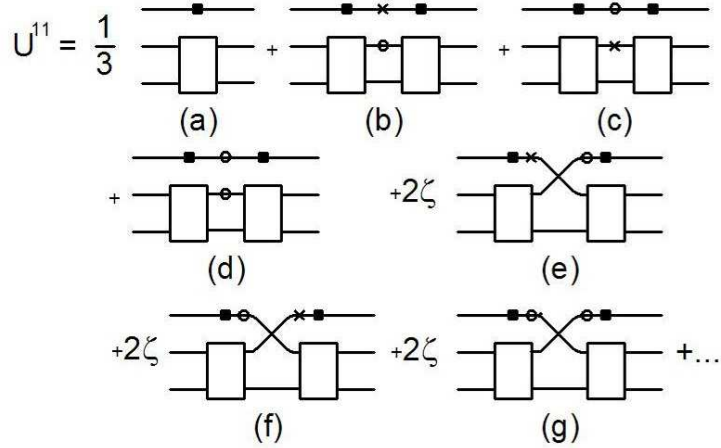}}
\caption{Contributions to $U^{11}$ up to second order in
Bethe-Salpeter kernel. The rectangular boxes represent the (symmetrized)
two-body Bethe-Salpeter kernel, the crosses represent on-shell propagators,
the open circles represent the off-shell propagators and the
filled squares represent factors of the inverse one-body
propagator.}\label{FigC}
\end{figure}

The lower right-hand contribution $U^{11}$ will also be
complicated by this mixing. Figure \ref{FigC} shows diagrams for
$U^{11}$ associated with the first iteration of (\ref{uoff}):
\begin{equation}
U^{11} \simeq \frac{1}{3}\overline{V}^1 iG_1^{-1}-
\frac{1}{3}\overline{V}^1iG_1^{-1}\Delta g^1_2 (1+ 2\zeta{\cal P}_{12})
iG_1^{-1} \overline{V}^1 \, .
\end{equation}
The rectangular boxes represent the two-body Bethe-Salpeter kernel
for particles two and three, the cross represents the on-shell
propagator $iQ_1$, the open circle represents the off-shell
propagator $\Delta G_i$ and the square box represents the inverse
one-body propagator $iG_1^{-1}$. These diagrams therefore contain
some ill-defined products such as $Q_1G_1^{-1}$ that can be resolved
only when the quasipotential is used as part of some larger diagram.
The resolution of these products is then obtained by using the rules
developed in \cite{2Ncur} which are repeated in Appendix C for
convenience.

Note that in Fig.~\ref{FigC},  particle 1 is disconnected from
particles 2 and 3 in diagrams (a-d), while the diagrams (e-g) are
completely connected. In (\ref{tgr}), the quasipotential appears
only with the external legs for particles 1 and 2 constrained to
be on shell due to (\ref{rule1}) and (\ref{rule2}).  Using these,
the completely constrained quasipotential contains only diagrams
(a) and (b) which are disconnected and have the form of two-body
interactions. As we will see below, the quasipotential with only
the initial or final legs constrained can contribute to effective
current operator and diagrams (e) and (f) can appear in that role.
Diagrams (c), (d) and (g) will never contribute.

%
% Fig 4
\begin{figure}
\centerline{\includegraphics[height=3in]{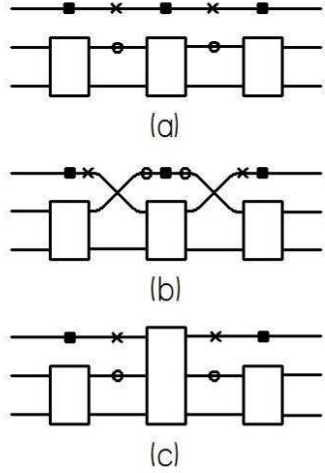}}
\caption{Selected diagrams contributing to $U^{11}$ at third order
in the Bethe-Salpeter kernels. The large rectangular box
represents the three-body Bethe-Salpeter kernel.}\label{FigD}
\end{figure}

There will be many diagrams that contribute to $U^{11}$ at the
second iteration of (\ref{uoff}). Examples of such diagrams are
shown in Fig. \ref{FigD}. Diagram (a) is disconnected and can be
categorized as a two-body interaction. Diagrams (b) and (c) are,
however, completely connected diagrams and are three-body
interactions. Therefore, $U^{11}$ contains both two- and
three-body interactions.

As a practical matter, it may be useful to separate the two- and
three-body interactions. This can be done by first defining a new,
diagonal quasipotential
\begin{equation}
\bm{w}=\bm{v}-\bm{v}\Delta
g^1_2\bm{\alpha}\bm{w}=\bm{v}-\bm{w}\Delta
g^1_2\bm{\alpha}\bm{v}=
\left(
\begin{array}{cc}
{\cal A}_3 W^{0} & 0\\
0 & \frac{1}{3} \overline{W}^{1} i G_1^{-1}
\end{array}
\right) \,, \label{wQP}
\end{equation}
where the diagonal matrix $\bm{\alpha}$ is given by
(\ref{Alfmat}). Thus the upper left-hand component of $\bm{w}$ is
an iteration of the three-body Bethe-Salpeter kernel, while the
lower right hand component is an iteration of the two-body
Bethe-Salpeter kernel:
\begin{equation}
\overline{W}^{1} = \overline{V}^{1}-
\overline{V}^{1}(-i\Delta G_2 G_3)\overline{W}^{1} \, ,
\end{equation}
i.e., $\overline{W}^{1}$ is the two-body spectator quasipotential for
particles 2 and 3.The component $W^{11}=\frac{1}{3} \overline{W}^{1}
i G_1^{-1}$ is identical to the two-body contribution to $U^{11}$.
Indeed, $\bm{u}$ is related to $\bm{w}$ by
\begin{equation}
\bm{u}=\bm{w}-\bm{w}\Delta g^1_2\bm{\beta}\bm{u}\,,\label{ufromw}
\end{equation}
where the matrix $\bm{\beta}$ is given by (\ref{beta}). The matrix
$\bm{w}$ can now be written as the sum of two- and three-body
contributions as
\begin{equation}
\bm{w}=\bm{w}^{(2)}+\bm{w}^{(3)}.\label{w23}
\end{equation}
Since $\bm{w}^{(2)}$ is identical to the two-body content of
$\bm{u}$,  $\bm{u}$ can be written as
\begin{equation}
\bm{u}=\bm{w}^{(2)}+\bm{u}^{(3)}.\label{u23}
\end{equation}
Substituting (\ref{w23}) and (\ref{u23}) into (\ref{ufromw}), the
equation for the three-body contributions to $\bm{u}$ is
\begin{equation}
\bm{u}^{(3)}=\bm{w}^{(3)}-\bm{w}\Delta
g^1_2\bm{\beta}\bm{w}^{(2)}-\bm{w}\Delta
g^1_2\bm{\beta}\bm{u}^{(3)}\,.\label{u3}
\end{equation}

Equation (\ref{toff}) can be resumed to give
\begin{equation}
\bm{t}=\bm{m}-\bm{m}{\cal Q}_1{\cal Q}_2 g^1_2\, \bm{\beta}\,
\bm{t}\, ,\label{tgr2}
\end{equation}
where
\begin{equation}
\bm{m}=\bm{u}-\bm{u}{\cal Q}_1{\cal Q}_2
g^1_2\bm{\alpha}\bm{m}\,.\label{mgr}
\end{equation}
Since, in general, $\bm{u}$ is not diagonal, $\bm{m}$ is also not
diagonal and all of its components contain three-body
interactions. It is also useful to separate the two- and
three-body contributions to $\bm{m}$. The two-body contribution
will just contain iterations of the two-body contribution to
$\bm{u}$. We can then define the two-body scattering matrix as
\begin{equation}
\bm{m}^{(2)}=\bm{w}^{(2)}-\bm{w}^{(2)}{\cal Q}_1{\cal Q}_2
g^1_2\bm{\alpha}\bm{m}^{(2)}\,.\label{m2}
\end{equation}
Substituting $\bm{m}=\bm{m}^{(2)}+\bm{m}^{(3)}$ into (\ref{mgr})
and using (\ref{m2}) the three-body scattering matrix is given by
\begin{equation}
\bm{m}^{(3)}=\bm{u}^{(3)}-\bm{u}^{(3)} {\cal Q}_1{\cal Q}_2
g^1_2\bm{\alpha}\bm{m}^{(2)}-\bm{u} {\cal Q}_1{\cal Q}_2
g^1_2\bm{\alpha}\bm{m}^{(3)}\,.
\end{equation}

As in the matrix form of BS equation (cf eqs.\
(\ref{GR2sym},\ref{GL2sym})), in the Gross formalism there are two
propagators, a right-hand one
\begin{equation}
\bm{g}^1_{2R}=g^1_2\bm{1}-g^1_2\bm{t}^1_2\bm{\delta}g^1_2
=g^1_2\bm{1}-g^1_2\bm{u}^1_2\bm{\delta}\bm{g}^1_{2R}\label{ggrR}
\end{equation}
with the inverse
\begin{equation}
({\bm{g}^1_{2R}})^{-1}=({g^1_2})^{-1}\bm{1}+
\bm{u}^1_2\bm{\delta}\,,\label{ggrRinv}
\end{equation}
and a left-hand one
\begin{equation}
\bm{g}^1_{2L}=g^1_2\bm{1}-g^1_2\bm{\delta}\bm{t}^1_2g^1_2
=g^1_2\bm{1}-\bm{g}^1_{2L}\bm{\delta}\bm{u}^1_2g^1_2\label{ggrL}
\end{equation}
with the inverse
\begin{equation}
({\bm{g}^1_{2L}})^{-1}=({g^1_2})^{-1}\bm{1}+\bm{\delta}\bm{u}^1_2\,.
\label{ggrLinv}
\end{equation}

The three-body bound-state vertex function can be obtained using
(\ref{tgr}) and the analytic form of the t-matrix at the
bound-state pole. The vertex function satisfies the equation
\begin{equation}
\left|\bm{\gamma}^1_2\right>=-\bm{u}^1_2\bm{\delta}g^1_2
\left|\bm{\gamma}^1_2\right>=
-\bm{m}^1_2\bm{\beta}g^1_2
\left|\bm{\gamma}^1_2\right>
\, .
\label{vertexgr}
\end{equation}
From the non-linear form of the t-matrix equation in the Gross
formalism
\begin{equation}
\bm{t}^1_2=\bm{u}^1_2-\bm{t}^1_2\bm{\delta}g^1_2\bm{t}^1_2-
\bm{t}^1_2g^1_2\bm{\delta}\bm{u}^1_2\bm{\delta}g^1_2\bm{t}^1_2
\end{equation}
we get the normalization condition for vertex function (\ref{vertexgr})
\begin{equation}
1=\left<\bm{\gamma}^1_2\right|\left[\frac{\partial g^1_2}{\partial
P^2}\bm{\delta}-g^1_2\bm{\delta}\frac{\partial\bm{u}^1_2}{\partial
P^2}\bm{\delta}g^1_2\right]_{P^2=M^2}\left|\bm{\gamma}^1_2\right>\,.
\end{equation}
The bound-state Gross wave function is defined as
\begin{equation}
\left|\bm{\psi}^1_{2}\right>=g^1_2\left|\bm{\gamma}^1_2\right>\,.
\end{equation}
Using (\ref{vertexgr}) it is straightforward to show that
\begin{equation}
({\bm{g}^1_{2R}})^{-1}\left|\bm{\psi}^1_{2}\right>=
\left<\bm{\psi}^1_{2}\right|({\bm{g}^1_{2L}})^{-1}=0\,.
\label{wfeqgr}
\end{equation}

The Gross wave functions for the scattering states can be obtained
by considering the analytic properties of
$({\bm{g}^1_{2L}})^{-1}$, and then the effective current operator
can  be determined in a similar fashion from the seven-point
function (\ref{7point2d}) and the substitution (\ref{grsub}). An
alternative approach is to express the Bethe-Salpeter wave
functions in terms of the corresponding Gross wave functions. To
do this substitute (\ref{grsub}) into the M\"oller operator giving
\begin{equation}
\bm{{1}}- \bm{\delta}\bm{ t}{G}^0_{BS}=\bm{{1}}- \bm{\delta}\bm{
t}\left(g^1_2{\cal Q}_1 {\cal Q}_2+\Delta g^1_2\right)=\bm{{1}}-
\bm{\delta}\bm{ t}g^1_2{\cal Q}_1 {\cal Q}_2-\bm{\delta}\bm{
t}\Delta g^1_2\,.
\end{equation}
Using (\ref{toff}) to replace the t-matrix in the last term gives
\begin{eqnarray}
\bm{{1}}-\bm{\delta}\bm{t}{G}^0_{BS}&=&\bm{{1}}- \bm{\delta}\bm{
t}g^1_2{\cal Q}_1 {\cal
Q}_2-\bm{\delta}\left(\bm{u}-\bm{t}g^1_2{\cal Q}_1 {\cal
Q}_2\bm{\delta}\bm{u}\right)\Delta g^1_2\nonumber\\
&=&\left(\bm{1}-\bm{\delta}\bm{t}g^1_2{\cal Q}_1 {\cal
Q}_2\right)\left(\bm{1}-\bm{\delta}\bm{u}\Delta g^1_2\right)\,.
\end{eqnarray}
Similarly, the right-hand M\"oller operator can be rewritten as
\begin{equation}
\bm{{1}}-{G}^0_{BS}\bm{t}\bm{\delta}=\left(\bm{1}-\Delta
g^1_2\bm{u}\bm{\delta}\right)\left(\bm{1}-g^1_2{\cal Q}_1 {\cal
Q}_2\bm{t}\bm{\delta}\right)\,.
\end{equation}
By examining the bound-state pole in the t-matrices, we can find
expressions for the Bethe-Salpeter bound-state wave functions in
term of the corresponding Gross wave functions:
\begin{equation}
\left<\bm{\psi}\right|=\left<\bm{\psi}^1_{2}\right|\left(\bm{1}-
\bm{\delta}\bm{u}\Delta g^1_2\right) \label{wffactL}
\end{equation}
and
\begin{equation}
\left|\bm{\psi}\right>=\left(\bm{1}-\Delta
g^1_2\bm{u}\bm{\delta}\right)\left|\bm{\psi}^1_{2}\right>\,.
\label{wffactR}
\end{equation}

The three-body scattering final state can be obtained directly
from the two-dimensional form of the corresponding Bethe-Salpeter
state (\ref{psi3sc2}) as
\begin{eqnarray}
\left<\bm{{\psi}}^{(-)} \right|&=&\frac{1}{4}
\left<\bm{p}_1,s_1;\bm{p}_2,s_2;\bm{p}_3,s_3\right| {\cal
A}_3\bm{{d}}^T\left[\bm{{1}}- \bm{\delta}
\bm{t} G^0_{BS}\right]\nonumber\\
&=&\frac{1}{4} \left<\bm{p}_1,s_1;\bm{p}_2,s_2;\bm{p}_3,s_3\right|
{\cal A}_3\bm{{d}}^T\left(\bm{1}-\bm{\delta}\bm{t}g^1_2{\cal Q}_1
{\cal
Q}_2\right)\left(\bm{1}-\bm{\delta}\bm{u}\Delta g^1_2\right)\nonumber\\
&=&\left<\bm{{\psi}}^{1(-)}_{2}\right|\left(\bm{1}-\bm{\delta}\bm{u}\Delta
g^1_2\right)
\end{eqnarray}
where
\begin{equation}
\left<\bm{{\psi}}^{1(-)}_{2}\right|= \frac{1}{4}
\left<\bm{p}_1,s_1;\bm{p}_2,s_2;\bm{p}_3,s_3\right| {\cal
A}_3\bm{{d}}^T\left[\bm{{1}}- \bm{\delta} \bm{t}^1_2 g^1_2\right]
\label{3wavegr}
\end{equation}
is the Gross three-body scattering state which satisfies the wave
equation
\begin{equation}
\left<\bm{{\psi}}^{1(-)}_{2}\right| ({\bm{g}^1_{2L}})^{-1}=0\,.
\end{equation}

Finding a similar relation for the two-body final scattering state
is a little more complicated since it contains a two-body bound
state which also needs to be re-expressed in terms of the
corresponding spectator wave function for two-body bound state as
well as rewriting a M\"oller operator of a slightly different form
than that examined above. The two-body final state in the
two-dimensional matrix form is given by (\ref{phim2d})
\begin{equation}
\left<\bm{\phi}^{1(-)}\right|=
\frac{1}{3}\left<\Phi^{(2)1};\bm{p}_1,s_1\right| {\cal A}_{23}
\left(
\begin{array}{cc}
0 & 1
\end{array}
\right)
\left(\bm{1}-\bm{\beta}\bm{t}\bm{G}^0_{BS}\right)\,.
\label{scat2bbs}
\end{equation}
The symmetrized two-body spectator bound-state  wave function is
for particles 2 and 3 is defined as
\begin{equation}
\left|\varphi^{(2)}_2\right>={\cal Q}_2{\cal
A}_{23}\left|\Phi^{(2)}\right>\,.
\end{equation}
and satisfies the equation
\begin{equation}
(G_3^{-1}+ \overline{W}^{11}_{22})\left| \varphi^{(2)}_{2} \right>
=0 \, ,
\end{equation}
where $\overline{W}^{11}_{22}={\cal Q}_2\overline{W}^1{\cal
Q}_1{\cal Q}_2$.
The corresponding Bethe-Salpeter wave function can be expressed in
terms of the spectator wave function one by
\begin{equation}
{\cal A}_{23}\left|\Phi^{(2)}\right> = \left[ 1- (-i \Delta G_2 G_3)
\overline{W}^1 \right] \left| \varphi^{(2)}_{2} \right> \, .
\end{equation}
The asymptotic spectator state satisfies the equation
\begin{equation}
\left<\varphi^{(2)}_{2};\bm{p}_1,s_1\right| \left[ G_3^{-1} +
\overline{W}^{11}_{22} \right] =  0 \, ,
\end{equation}
and the asymptotic state in (\ref{scat2bbs}) can be now written in
terms of this spectator state as
\begin{equation}
\left<\Phi^{(2)1};\bm{p}_1,s_1\right|{\cal A}_{23}=
\left<\varphi^{(2)}_{2};\bm{p}_1,s_1\right|{\cal Q}_1\left[1-
\overline{W}^{1}\left(-i{\cal Q}_1\Delta G_2G_3\right)\right] \, .
\end{equation}
From ${\cal Q}_1 \Delta g^1_2= -i{\cal Q}_1 \Delta G_2 G_3$, it
follows
\begin{equation}
{\cal Q}_1\, \left[1- \overline{W}^{1}\left(-i{\cal Q}_1\Delta G_2
G_3\right)\right] \left(
\begin{array}{cc}
0 & 1
\end{array}
\right) = {\cal Q}_1\, \left(\begin{array}{cc} 0 & 1
\end{array}
\right)\left(\bm{1}-\bm{\alpha}\bm{w}\Delta g^1_2\right)\,.
\end{equation}
The Bethe-Salpeter scattering state can now be written as
\begin{eqnarray}
\left<\bm{\phi}^{1(-)}\right|&=&
\frac{1}{3}\left<\varphi^{(2)}_{2};\bm{p}_1,s_1\right| \left(
\begin{array}{cc}
0 & 1
\end{array}
\right)\left(\bm{1}-\bm{\alpha}\bm{w}\Delta g^1_2\right)
\left(\bm{1}-\bm{\beta}\bm{t}\bm{G}^0_{BS}\right)
\end{eqnarray}
Using the identity
\begin{equation}
\left(\bm{1}-\bm{\alpha}\bm{w}\Delta g^1_2\right)
\left(\bm{1}-\bm{\beta}\bm{t}\bm{G}^0_{BS}\right)
=\left[\bm{1}-\left(\bm{1}-\bm{\alpha}\bm{w}\Delta
g^1_2\right)\bm{\beta}\bm{t}{\cal Q}_1 {\cal Q}_2 g^1_2\right]
\left(\bm{1}-\bm{\delta}\bm{u}\Delta g^1_2\right)\,,
\end{equation}
this can be rewritten as
\begin{eqnarray}
\left<\bm{\phi}^{1(-)}\right|&=&
\frac{1}{3}\left<\varphi^{(2)}_{2};\bm{p}_1,s_1\right| \left(
\begin{array}{cc}
0 & 1
\end{array}
\right) \left[\bm{1}-\left(\bm{1}-\bm{\alpha}\bm{w}\Delta
g^1_2\right)\bm{\beta}\bm{t}{\cal Q}_1 {\cal Q}_2 g^1_2\right]
\left(\bm{1}-\bm{\delta}\bm{u}\Delta g^1_2\right)\nonumber\\
&=&\left<\bm{\phi}^{1(-)}_{2}\right|\left(\bm{1}-\bm{\delta}\bm{u}\Delta
g^1_2\right)\,,
\end{eqnarray}
where
\begin{equation}
\left<\bm{\phi}^{1(-)}_{2}\right|=\frac{1}{3}
\left<\varphi^{(2)}_{2};\bm{p}_1,s_1\right| \left(
\begin{array}{cc}
0 & 1
\end{array}
\right) \left[\bm{1}-\left(\bm{1}-\bm{\alpha}\bm{w}\Delta
g^1_2\right)\bm{\beta}\bm{t}{\cal Q}_1 {\cal Q}_2 g^1_2\right]
\label{2wavegr}
\end{equation}
is the Gross two-body scattering state. It satisfies the equation
\begin{equation}
\left<\bm{\phi}^{1(-)}_{2}\right| ({\bm{g}^1_{2L}})^{-1}=0\,.
\end{equation}

\subsection{Electromagnetic current}

According to previous subsection,  the relationship of the
Bethe-Salpeter to Gross wave functions always involves a factor of
$\left(\bm{1}-\Delta g^1_2\bm{u}\bm{\delta}\right)$ for initial
states and $\left(\bm{1}-\bm{\delta}\bm{u}\Delta g^1_2\right)$ for
final states. Any Bethe-Salpeter current matrix element can now be
converted to the corresponding Gross matrix element by defining the
effective current for the Gross equation as
\begin{equation}
\bm{j}^{1\mu}_{2\,{\rm eff}}={\cal Q}_1{\cal Q}_2
\left(\bm{1}-\bm{\delta}\bm{u}\Delta g^1_2\right)\bm{j}^{\mu}_{\rm eff}
\left(\bm{1}-\Delta g^1_2\bm{u}\bm{\delta}\right){\cal Q}_1{\cal Q}_2\,,
\label{jgr}
\end{equation}
where $\bm{j}^{\mu}_{\rm eff}$ is given by (\ref{jbs2d}) and
(\ref{jbsint2d}).

Contraction of the photon four-momentum with this effective
current operator yields
\begin{eqnarray}
q_\mu \bm{j}^{1\mu}_{2\,{\rm eff}}&=&{\cal Q}_1{\cal
Q}_2\left(\bm{1}-\bm{\delta}\bm{u}\Delta
g^1_2\right)\left(e_T(q)\bm{\delta}\bm{g}_R^{-1}
-\bm{g}_L^{-1}e_T(q)\bm{\delta}\right)\left(\bm{1}-\Delta
g^1_2\bm{u}\bm{\delta}\right){\cal Q}_1{\cal Q}_2\nonumber\\
&=&{\cal Q}_1{\cal
Q}_2\left[e_3(q)\bm{\delta}\left(G_3^{-1}\bm{1}+\bm{u}\bm{\delta}\right)
-\left(G_3^{-1}\bm{1}+\bm{\delta}\bm{u}\right)e_3(q)\bm{\delta}\right]
{\cal Q}_1{\cal Q}_2\nonumber\\
&=&e_3(q)\bm{\delta}{\cal Q}_1{\cal Q}_2 ({\bm{g}^1_{2R}})^{-1}
-({\bm{g}^1_{2L}})^{-1}{\cal Q}_1{\cal Q}_2\bm{\delta}e_3(q) \, ,
\end{eqnarray}
where the rules of Appendix C have been used to simplify this
expression (the proof is similar to that for a two-body system
described in detail in Sec.~3.2 of ref.~\cite{2Ncur}). This, along
with the wave equations for the Gross wave functions, implies that
the current is conserved.

To bring the expression for the effective current to a form that can
be more readily used for practical calculation requires use of the
non-associative algebra for the on-shell operators ${\cal Q}_i$ and
the propagators developed in \cite{2Ncur} and summarized in Appendix
C. The pieces of primary concern are those involving the one-body
current and can be simplified with the following identities:
\begin{equation}
{\cal Q}_1{\cal Q}_2J^{(1)\mu}{\cal Q}_1{\cal Q}_2= {\cal
Q}_1{\cal Q}_2J^\mu_3\,,\label{ja}
\end{equation}
\begin{equation}
{\cal Q}_1{\cal Q}_2J^{(1)\mu}\Delta g^1_2= {\cal Q}_1{\cal
Q}_2\left( J^\mu_1 G_1+J^\mu_2 G_2 \right)\,, \label{jb}
\end{equation}
\begin{equation}
\Delta g^1_2J^{(1)\mu}{\cal Q}_1{\cal Q}_2= \left( G_1 J^\mu_1 +
G_2 J^\mu_2 \right){\cal Q}_1{\cal Q}_2 \label{jc}
\end{equation}
and
\begin{eqnarray}
\Delta g^1_2J^{(1)\mu}\Delta g^1_2&=&
- \left(i{\cal Q}_1 J^\mu_1 G_1+ G_1 J^\mu_1 i{\cal Q}_1\right)
\Delta G_2 G_3 - (\Delta G_1 J^\mu_1 \Delta G_1)G_2 G_3 \nonumber\\
&&- \Delta G_1 \left(i{\cal Q}_2 J^\mu_2 G_2+ G_2 J^\mu_2 i{\cal Q}_2\right)G_3 -
G_1(\Delta G_2 J^\mu_2 \Delta G_2) G_3 \nonumber\\
&&+ \left( -i{\cal Q}_1 \Delta G_2- \Delta G_1 i{\cal Q}_2- \Delta G_1 \Delta G_2 \right)
(G_3 J^\mu_3 G_3) \, .
\label{jd}
\end{eqnarray}
In these expressions we have used the identity ${\cal Q}_iJ_i{\cal
Q}_i=0$ (\ref{rule6}) to give ${\cal Q}_iJ_i\Delta G_i={\cal
Q}_iJ_i G_i$ and $\Delta G_iJ_i{\cal Q}_i= G_iJ_i{\cal Q}_i$. The
corresponding part of the effective spectator current then reads:
\begin{eqnarray}
\bm{j}^{1\mu}_{2\,{\rm eff,IA}}&=& {\cal Q}_1{\cal Q}_2\, \left[
J^\mu_3  \bm{\delta} -\bm{\delta}\bm{u}(G_1 J^\mu_1 + G_2
J^\mu_2)\bm{\delta}
- \bm{\delta} ( J^\mu_1 G_1+J^\mu_2 G_2)\bm{u} \bm{\delta} \right. \nonumber\\
&&   \left. + \bm{\delta} \bm{u} (\Delta g^1_2J^{(1)\mu}\Delta g^1_2)
\bm{\delta} \bm{u}\bm{\delta} \right]\, {\cal Q}_1{\cal Q}_2 \, ,
\label{jspecIA}
\end{eqnarray}
where $\Delta g^1_2J^{(1)\mu}\Delta g^1_2$ from the last term is
given by eq.~(\ref{jd}).

The interaction current also contains a piece containing the
one-body current operator that resulted from our symmetric
separation of the seven-point function into a current operator and
six-point propagators:
\begin{equation}
\bm{j}^\mu_{sym}=\frac{1}{3}\left(\begin{array}{c} 0 \\
1\end{array}\right)i\overline{V}^1J^\mu_1\left(\begin{array}{cc}
0&1
\end{array}\right)\,.
\end{equation}
Contributions from this current can be simplified using the
identities:
\begin{equation}
Q_1Q_2 i\overline{V}^1 J^\mu_1Q_1Q_2=0\, ,
\label{je}
\end{equation}
\begin{equation}
Q_1Q_2i\overline{V}^1J^\mu_1\Delta g^1_2
= - (Q_1 J^\mu_1 G_1) (Q_2 i\overline{V}^1  G_2 G_3) \, ,
\label{jf}
\end{equation}
\begin{equation}
\Delta g^1_2i\overline{V}^1J^\mu_1Q_1Q_2 =
- (G_1 J^\mu_1 Q_1 )( G_2 G_3 i\overline{V}^1 Q_2) \, ,
\label{jg}
\end{equation}
and
\begin{eqnarray}
\Delta g^1_2i\overline{V}^1J^\mu_1\Delta g^1_2&=&
(iQ_1 J^\mu_1 G_1)(\Delta G_2 G_3 i\overline{V}^1  G_2 G_3)
+ (G_1 J^\mu_1 iQ_1) (G_2 G_3 i\overline{V}^1\Delta G_2 G_3)\nonumber\\
&&
+ (\Delta G_1 J^\mu_1 \Delta G_1) (G_2 G_3 i\overline{V}^1 G_2 G_3)
\,.
\label{jh}
\end{eqnarray}
The corresponding effective spectator current is given by:
\begin{eqnarray}
\bm{j}^{1\mu}_{2\,{\rm eff,sym}}&=& {\cal Q}_1{\cal Q}_2\,
\frac{1}{3}\, \bm{\delta}\, \left[ \bm{u}\bm{\delta}
\left(
\begin{array}{cc}
0 & 0 \\
0 & G_1 J^\mu_1 i\overline{V}^1 G_2 G_3
\end{array}
\right) +
\left(
\begin{array}{cc}
0 & 0 \\
0 & J^\mu_1 G_1 i\overline{V}^1 G_2 G_3
\end{array}
\right)\bm{\delta}\bm{u} \right. \nonumber\\
&& \left. + \bm{u}\bm{\delta}
\left(
\begin{array}{cc}
0 & 0 \\
0 & \Delta g^1_2i\overline{V}^1J^\mu_1\Delta g^1_2
\end{array}
\right)\bm{\delta}\bm{u} \right]\, \bm{\delta}\, {\cal Q}_1{\cal Q}_2 \, .
\label{jspecsym}
\end{eqnarray}

The two-body exchange current part of the interaction current is
\begin{equation}
\bm{j}^{(2)\mu}=\frac{1}{3}\left(\begin{array}{c} 0 \\
1\end{array}\right) \overline{J}^{1\mu}_{ex} \left(\begin{array}{cc}
0&1
\end{array}\right)\,,
\end{equation}
where $\overline{J}^{1\mu}_{ex}= iG_1^{-1}\overline{J}^{(2)\mu}_{23}$.
Contributions from this current can be simplified using the identities
\begin{equation}
Q_1Q_2\, \overline{J}^{1\mu}_{ex}\,  Q_1Q_2=
Q_1Q_2\overline{J}^{(2)\mu}_{23}Q_2\,,
\label{ji}
\end{equation}
\begin{equation}
Q_1Q_2\, \overline{J}^{1\mu}_{ex} \Delta g^1_2=
-iQ_1Q_2\overline{J}^{(2)\mu}_{23}\Delta G_2G_3\,,
\label{jj}
\end{equation}
\begin{equation}
\Delta g^1_2\,  \overline{J}^{1\mu}_{ex}\, Q_1Q_2=
-iQ_1\Delta G_2G_3\overline{J}^{(2)\mu}_{23}Q_2
\label{jk}
\end{equation}
and
\begin{equation}
\Delta g^1_2\, \overline{J}^{1\mu}_{ex}\, \Delta g^1_2=
-Q_1\Delta G_2G_3 \overline{J}^{(2)\mu}_{23}\Delta G_2 G_3 +
i\Delta G_1 G_2 G_3 \overline{J}^{(2)\mu}_{23} G_2 G_3\,.
\label{jl}
\end{equation}
The spectator two-body exchange current is thus
\begin{eqnarray}
\bm{j}^{1\mu}_{2\,{\rm eff,ex2}}&=& {\cal Q}_1{\cal Q}_2\,
\frac{1}{3}\, \bm{\delta}\, \left[
\left(
\begin{array}{cc}
0 & 0 \\
0 & \overline{J}^{(2)\mu}_{23}
\end{array}
\right) +
\bm{u}\bm{\delta}
\left(
\begin{array}{cc}
0 & 0 \\
0 &  \Delta G_2 G_3 i \overline{J}^{(2)\mu}_{23}
\end{array}
\right) +
\left(
\begin{array}{cc}
0 & 0 \\
0 & i \overline{J}^{(2)\mu}_{23} \Delta G_2 G_3
\end{array}
\right)\bm{\delta}\bm{u} \right. \nonumber\\
&& \left. + \bm{u}\bm{\delta}
\left(
\begin{array}{cc}
0 & 0 \\
0 & \Delta g^1_2 \overline{J}^{1\mu}_{ex} \Delta g^1_2
\end{array}
\right)\bm{\delta}\bm{u} \right]\, \bm{\delta}\, {\cal Q}_1{\cal Q}_2 \, .
\label{jspecex2}
\end{eqnarray}

Since the Bethe-Salpeter three-body current is completely
connected there is no simplification to the contributions from
this current:
\begin{equation}
\bm{j}^{1\mu}_{2\,{\rm eff,ex3}} = {\cal Q}_1{\cal Q}_2
\left(\bm{1}-\bm{\delta}\bm{u}\Delta g^1_2\right)\,
\bm{\delta}\,
\left(
\begin{array}{cc}
\overline{J}^{0\mu}_{ex} & 0 \\
0 & 0
\end{array}
\right)
\bm{\delta}\,
\left(\bm{1}-\Delta g^1_2\bm{u}\bm{\delta}\right){\cal Q}_1{\cal Q}_2\,,
\label{jspecex3}
\end{equation}

The full effective spectator current is the sum of these four
contributions (\ref{jspecIA},\ref{jspecsym},\ref{jspecex2},\ref{jspecex3}):
\begin{equation}
\bm{j}^{1\mu}_{2\,{\rm eff}} = \bm{j}^{1\mu}_{2\,{\rm eff,IA}} +
\bm{j}^{1\mu}_{2\,{\rm eff,sym}} +\bm{j}^{1\mu}_{2\,{\rm eff,ex2}} +
\bm{j}^{1\mu}_{2\,{\rm eff,ex3}} \, .
\label{jspecful}
\end{equation}

Clearly, this effective current is quite complicated in its
complete form and would correspond to a large number of Feynman
diagrams. For the purpose of comparing these result to those of
\cite{kb97} and \cite{Gr04} it is useful to compare the simplest
case where the interaction is given by a one-boson-exchange kernel
and no three-body contributions are present.

%___________________________________________________________________
\section{Spectator formalism in the one-boson-exchange approximation}

The expressions presented above for the spectator formalism are
exact. However, in any practical calculation using either the
Bethe-Salpeter or spectator approaches it is necessary to approximate
the equations by truncating the kernels. This is usually done by
expanding the kernels in the number of meson exchanges. Indeed, the
only solutions that have been obtained for the three-body spectator
equations at this time are in the one-boson-exchange approximation.
Care must be taken in truncating the kernel in order to maintain the
wave equations and Ward identities. In this section, we will simplify
the general expressions to the case of the one-boson-exchange (OBE)
approximation.

In the OBE approximation the quasipotentials $\bm{u}$ and $\bm{w}$
are equal to Bethe-Salpeter kernel $\bm{v}$ and there are no
three-body forces or currents. Then the two-channel formalism
developed above reduces to just one independent channel: there is
only one independent component of the t-matrix, of the effective
current and of the wave functions.

The general expressions for the wave functions and currents given in
the previous section can  be reduced to this approximation by
replacing $\bm{u}$ and $\bm{w}$ by
\begin{equation}
\bm{v}=\left(
\begin{array}{cc}
0 & 0 \\
0 & \frac{1}{3}\overline{V}^{1}iG^{-1}_1
\end{array}
\right)\,.
\end{equation}
This implies that the spectator t-matrix becomes
\begin{equation}
\bm{t}^1_2=\left(
\begin{array}{cc}
0 & 0 \\
0 & T^{11}_{22}
\end{array}
\right)\,,
\label{toneb}
\end{equation}
where, using (\ref{tgr}) with the shorthand notation ${\cal P}= 1+
2\zeta {\cal P}_{12}$ and $\overline{V}^{11}_{22}= {\cal Q}_1{\cal
Q}_2 \overline{V}^{1}{\cal Q}_1{\cal Q}_2$, the nonzero component
of $\bm{t}$ is given by \cite{Gr82}
\begin{equation}
 T^{11}_{22}= \frac{1}{3}\overline{V}^{11}_{22}-
\overline{V}^{11}_{22}\, G_3\, {\cal P}\, T^{11}_{22}=
 \frac{1}{3}\overline{V}^{11}_{22}-T^{11}_{22}\, G_3\, {\cal P}\,
\overline{V}^{11}_{22} \, .
\label{tgrone}
\end{equation}
The propagators  (\ref{ggrR}) and (\ref{ggrL}), and their inverses
(\ref{ggrRinv}) and (\ref{ggrLinv}) in the subspace of this
symmetrized Faddeev component read:
\begin{eqnarray}
g^1_{2R}= G_3- 3G_3\,  T^{11}_{22}\, {\cal P}\, G_3 \, , &&
(g^1_{2R})^{-1}= G_3^{-1}+ \overline{V}^{11}_{22}\, {\cal P} \, , \\
g^1_{2L}= G_3- 3G_3{\cal P}\,  T^{11}_{22}\, G_3 \, , &&
(g^1_{2L})^{-1}= G_3^{-1}+ {\cal P}\, \overline{V}^{11}_{22} \, .
\label{ggroneL}
\end{eqnarray}

The effective current, defined by
(\ref{jspecIA},\ref{jspecsym},\ref{jspecex2},\ref{jspecex3},\ref{jspecful}),
can be simplified by setting the three-body Bethe-Salpeter current to
zero, and keeping only terms involving a single boson exchange. This
means that in (\ref{jspecIA}), only terms up to first order in
$\bm{u}=\bm{v}$ can be retained. In (\ref{jspecsym}) and
(\ref{jspecex2}), no terms containing $\bm{u}$ can be retained since
the corresponding Bethe-Salpeter currents already contain OBE
contributions. Since the three-body current is set to zero,
$\bm{j}^{1\mu}_{2\,{\rm eff,ex3}} = 0$. Using the rules from Appendix
C, the effective current in the OBE approximation is then (still in
the two-channel notation):
\begin{equation}
\bm{j}^{1\mu}_{2\,{\rm eff}}={\cal Q}_1{\cal Q}_2\left[
J^\mu_3\bm{\delta}-\bm{\delta}\left(
\begin{array}{cc}
0 & 0 \\
0 &\frac{1}{3}\left(
\overline{V}^1 G_2 J^\mu_2 + J^\mu_2 G_2 \overline{V}^1-\overline{J}^{(2)}_{23}
\right)
\end{array}
\right)\bm{\delta}\right]{\cal Q}_1{\cal Q}_2 \, .
\label{jeffoneb}
\end{equation}
To reduce it further to the effective operator in the one-channel
space of the OBE approximation, we have to specify first the
corresponding wave functions.

\begin{figure}
\centerline{\includegraphics[height=3in]{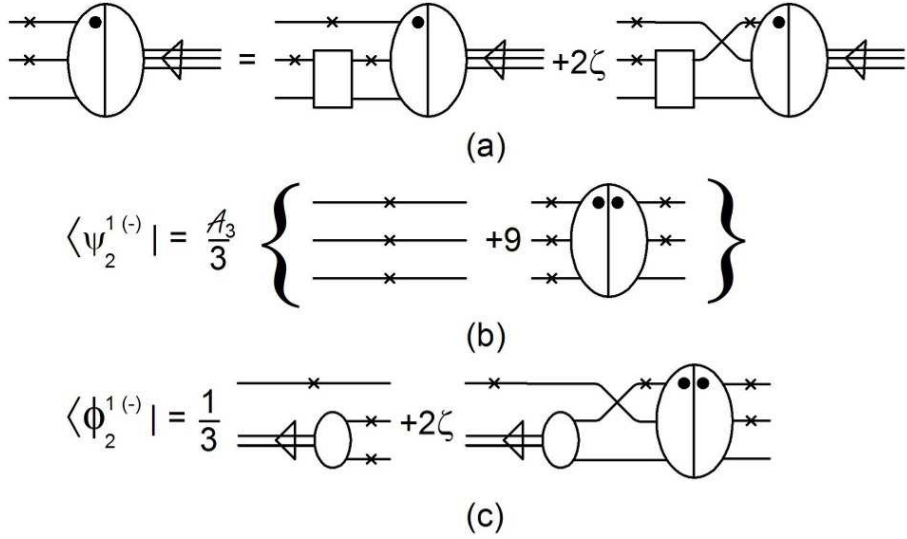}}
\caption{Feynman diagrams representing (a) the bound-state wave
equation, (b) the three-body final-state wave function and (c) the
two-body final-state wave function in the one-boson exchange
approximation.}\label{fig:waves}
\end{figure}

Since the three-body contributions have been eliminated, the
bound-state wave function (\ref{wfeqgr}) has only one nonzero
component,
\begin{equation}
\left|\bm{\psi}^1_2\right>\cong
\left(
\begin{array}{c}
0 \\
\left|\psi^1_2\right>
\end{array}
\right)
\end{equation}
where
\begin{equation}
(g^1_{2R})^{-1}\, \left|\psi^1_2 \right>= 0  \, .
\end{equation}
This wave equation is represented diagrammatically by
Fig.~\ref{fig:waves}a.
%
% Fig 5

The interaction-dependent part of the wave function
(\ref{3wavegr}) with the t-matrix (\ref{toneb}) already has only
one component, but the free propagation of three particles is
partially included in the 0$^{\rm th}$ channel, omitted now in the
OBE approximation. Notice that the current (\ref{jeffoneb}) always
acts on the wave functions both from the left-hand and right-hand
side with the matrix $\bm{\delta}$. This actually holds for any
properly symmetrized one- and two-body operator, the matrix
$\bm{\delta}$ just counts all symmetrization factors. Under the
action of $\bm{\delta}$ the three-body scattering wave function in
the OBE approximation can be replaced by its one component form
\begin{eqnarray}
\bm{\delta}\, \left| \bm{{\psi}}^{1(-)}_{2}\right> &=&
\bm{\delta}\, \frac{1}{4} \left(\bm{{1}}-  g^1_2 \bm{t}^1_2 \bm{\delta}
\right)\, \bm{d} {\cal A}_3
\left|\bm{p}_1,s_1;\bm{p}_2,s_2;\bm{p}_3,s_3\right> \nonumber\\
& \rightarrow &
\left(
\begin{array}{c}
1 \\
{\cal P}
\end{array}
\right)
{\cal A}_3\, (1- 9 G_3 T^{11}_{22} )\,
\left|\bm{p}_1,s_1;\bm{p}_2,s_2;\bm{p}_3,s_3\right>
=
\bm{\delta} \,
\left(
\begin{array}{c}
0 \\
\left|\psi^{1(-)}_{2}\right>
\end{array}
\right)
 \,  , \\
\left|\psi^{1(-)}_{2}\right> &=&
\frac{1}{3}
\left( 1- 9 G_3 T^{11}_{22}  \right)
{\cal A}_3\, \left|\bm{p}_1,s_1;\bm{p}_2,s_2;\bm{p}_3,s_3\right>
 \, .
\label{3wfgrone}
\end{eqnarray}
which follows from identities ${\cal A}_3{\cal P}=3{\cal A}_3$  and
\begin{eqnarray}
 \frac{1}{4} \bm{\delta}\, \bm{d}\, {\cal A}_3  &=& {\cal A}_3
\left(
\begin{array}{c}
1 \\
{\cal P}
\end{array}
\right)
=
\bm{\delta}\,
\left(
\begin{array}{c}
0 \\
\frac{1}{3}{\cal A}_3
\end{array}
\right) \, , \\
\bm{\delta}\,
\left(
\begin{array}{c}
0 \\
\left|\psi \right>
\end{array}
\right)&=&
3
\left(
\begin{array}{c}
1 \\
{\cal P}
\end{array}
\right)\, \left|\psi \right> \, .
\label{delpsi}
\end{eqnarray}
Also the three-body scattering wave function in the final state
multiplied by the matrix $\bm{\delta}$ is just
\begin{equation}
 \left< \bm{{\psi}}^{1(-)}_{2}\right|\, \bm{\delta}=
\left(0, \left<\psi^{1(-)}_{2}\right|\, \right) \, \bm{\delta} \, .
\label{3wfgronef}
\end{equation}
Thus in the considered OBE approximation the matrix elements of the
current (\ref{jeffoneb}) should be evaluated with the one-component
scattering wave functions (\ref{3wfgrone}). This wave function is
diagrammatically represented by Fig. \ref{fig:waves}b. From the
relations (\ref{tgrone},\ref{ggroneL}) the proper wave equation
\begin{equation}
\left<\psi^{1(-)}_{2}\right|\, (g^1_{2L})^{-1} = 0
\end{equation}
follows, which indicates that the one-component expression
(\ref{3wfgrone}) {\em is} the correct three-body scattering wave
function in the OBE approximation (i.e., not just effectively in the
matrix elements of operators proportional to symmetrizing matrix
$\bm{\delta}$). This can indeed be verified by repeating the whole
construction of this paper for the theory without the explicit
three-body force.

Some care must be used in truncating the final two-body scattering
state (\ref{2wavegr}). The factor $\bm{1}-\bm{\alpha}\bm{w}\Delta
g^1_2$ must also be expanded in order for this wave function to
satisfy the wave equation in the OBE approximation. In this the
factor $\bm{w}$ must be expanded to one order less than the order of
expansion. Since we are keeping only OBE terms, $\bm{w}$ should
contain zero boson exchanges which means that it should be set to
zero. This state then also simplifies to the one-component form
\begin{eqnarray}
\left< \bm{\phi}^{1(-)}_2 \right|&=& \left(0,\,
\left< \phi^{1(-)}_2 \right|\, \right) \, , \\
\left< \phi^{1(-)}_2 \right| &=& \frac{1}{3} \left<
\varphi^2_{Bs};\bm{p}_1,s_1\right|
 {\cal Q}_1 {\cal Q}_2
\left[1- 6\zeta {\cal P}_{12}T^{11}_{22} G_3 \right] \, , \\
\left< \phi^{1(-)}_2 \right|\, (g^1_{2L})^{-1}&=& 0 \, .
\end{eqnarray}
This wave function is represented diagrammatically by Fig.
\ref{fig:waves}c.

Since all wave function have in the OBE approximation only one
non-zero component, the matrix element of the effective current
(\ref{jeffoneb}) can be with the help of (\ref{delpsi}) rewritten as
a matrix element of the effective one-component current
$\left(J^\mu_{\rm eff} \right)^{11}_{22}$
\begin{eqnarray}
\left< \bm{\psi'} \right|\, \bm{j}^{1\mu}_{2\,{\rm eff}}\,
\left| \bm{\psi'} \right> &=&
\left< \psi' \right|\,  \left(J^\mu_{\rm eff}\right)^{11}_{22}\,
\left| \psi \right> \, , \\
\left( J^\mu_{\rm eff} \right)^{11}_{22}&=& {\cal Q}_1{\cal Q}_2\,
3\, \left[\, J^\mu_3{\cal P}- {\cal P}\left(\overline{V}^1 G_2 J^\mu_2+
J^\mu_2 G_2 \overline{V}^1- \overline{J}^{(2)}_{23}\right){\cal P}\,
\right]{\cal Q}_1{\cal Q}_2\, ,
\end{eqnarray}
where $\psi$ and $\psi'$ are any of the wave functions discussed above.
The Ward identity for this current is
\begin{eqnarray}
q_\mu \left( J^\mu_{\rm eff} \right)^{11}_{22}&=& 3 \left\{ \left[
e_3(q),G_3^{-1} \right] {\cal Q}_1 {\cal Q}_2 {\cal P} - {\cal P}
{\cal Q}_1 {\cal Q}_2 \left(\left[ e_2(q), \overline{V}^1 \right]-
\left[ e_2(q)+e_3(q), \overline{V}^1 \right]\right) {\cal Q}_1
{\cal Q}_2 {\cal P} \right\}\nonumber\\
&=& 3 \left\{ \left[ e_3(q),G_3^{-1} \right] {\cal Q}_1 {\cal Q}_2
{\cal P} +
{\cal P} \left[e_3(q), \overline{V}^{11}_{22}\right] {\cal P} \right\} \\
&=& 3 \left\{ {\cal Q}_1 {\cal Q}_2 {\cal P} e_3(q)
(g^1_{2R})^{-1} - (g^1_{2L})^{-1} e_3(q) {\cal Q}_1 {\cal Q}_2
{\cal P} \right\}\,. \label{Wardgr1}
\end{eqnarray}
This, along with the wave equations, shows that the current matrix
elements will be gauge invariant.

\begin{figure}
\centerline{\includegraphics[height=3in]{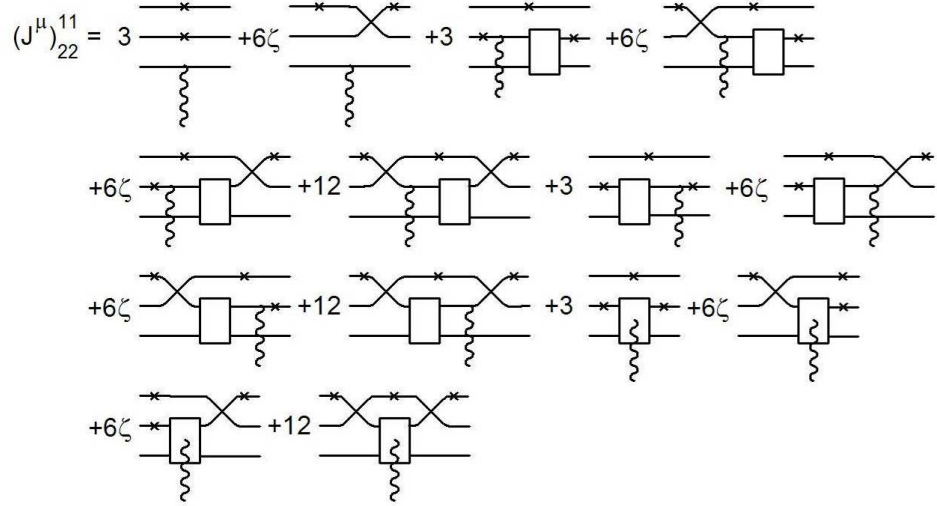}}
\caption{Feynman diagrams representing the effective current
operator for the spectator equation in the one-boson-exchange
approximation.}\label{fig:current}
\end{figure}

Figure \ref{fig:current} provides a diagrammatic representation of
all of the contributions to the one-boson-exchange Gross current.
%
% Fig 6
The matrix elements for various processes can now be produced by
combining the spectator wave functions with this effective
current. Some simplification can be obtained by defining off-shell
wave functions and scattering matrices. Doing this leads to expressions
coinciding with the results of \cite{kb97} and \cite{Gr04}.

\section{Summary}

In this paper, we have produced a comprehensive and general
description of the three-body spectator equations and the
corresponding electromagnetic currents starting with the
three-body Bethe-Salpeter equations and using an operator
formalism that has been developed for this purpose. Wave functions
for bound state, and two- and three-body states are presented
along with the effective electromagnetic current operator. The
results obtained are exact and can be truncated for practical
calculation. As an example we have shown truncation according to
the one-boson-exchange approximation is consistent with previous
results.

\acknowledgments

We are happy to acknowledge the support of the DOE through Jefferson
Laboratory, and  we gratefully acknowledge the support of the DOE
through grant No. DE-FG02-97ER41028 (for JWVO). JA was also supported
by grant GA CR 202/03/0210 and by the projects K1048102, ASCR
AV0Z1048901.

\appendix

\section{Matrix form of the BS equation: Distinguishable Particles}
%=========================Appendices=================================
%\newpage
%\renewcommand{\theequation}{A.\arabic{equation}}
%\setcounter{equation}{0}
%\section*{Appendix A: Matrix form of the BS equation: Distinguishable Particles}
%\setcounter{subsection}{0}
\subsection{Propagators and wave functions}
To make the equation for t-matrix, one can rewrite them by
defining the matrix components
\begin{eqnarray}
{\cal V}^{ij}&=&V^i iG^{-1}_i \delta_{ij}= {\cal V}^i \delta_{ij}
\, ,
\qquad i,j=1,2,3   \\
{\cal V}^{i0}&=&{\cal V}^{0i}=V^0\delta_{i0}= {\cal V}^0
\delta_{i0}\,  , \label{calVi}
\end{eqnarray}
and the same for matrices ${\cal M}$. In terms of these, the
equations for the three-body t-matrix reads
\begin{equation}
{\cal T}^{ij}={\cal V}^{ij}-\sum_{l=0}^3{\cal V}^{il}
G^0_{BS}\sum_{k=0}^3 {\cal T}^{kj} = {\cal
M}^{ij}-\sum_{l=0}^3{\cal M}^{il}G^0_{BS}\sum_{k\neq l} {\cal
T}^{kj}  \, , \label{tij-old}
\end{equation}
where $i, j$ now run from 0 to 3. We will often use the matrix
$\bf B$ with components
\begin{equation}
B_{ij}\equiv 1-\delta_{ij} \, ,
\end{equation}
which allows to re-write the equation for t-matrix in the form:
\begin{equation}
{\cal T}^{ij}={\cal M}^{ij}-\sum_{l,k=0}^3 {\cal
M}^{il}G^0_{BS}B_{lk} {\cal T}^{kj}  \, .
\end{equation}
The above form suggests a convenient matrix notation in the
``channel space'' defined by respective Faddeev components
$(\bm{{\cal T}})_{ij}= {\cal T}^{ij}, i,j= 0,1,2,3$. In terms of
such matrices
\begin{eqnarray}
\bm{{\cal T}}&=& \bm{{\cal M}}-\bm{{\cal M}}G^0_{BS}\bm{{B}}
\bm{{\cal T}}=\bm{{\cal M}}-\bm{{\cal T}}\bm{{B}}G^0_{BS}
\bm{{\cal M}}
\label{t_BS} \\
 &=& \bm{{\cal V}}-\bm{{\cal V}}G^0_{BS}(\bm{{1+B}}) \bm{{\cal
T}}=\bm{{\cal V}}-\bm{{\cal T}}(\bm{{1+B}})G^0_{BS}\bm{{\cal V}}
 \, ,
\label{t_BS2}
\end{eqnarray}
with
\begin{equation}
\bm{{\cal M}}=\bm{{\cal V}}-\bm{{\cal V}}G^0_{BS}\bm{{\cal M}}=
\bm{{\cal V}}-\bm{{\cal M}}G^0_{BS}\bm{{\cal V}} \, .
\label{m-matrix}
\end{equation}

The complete three-body t-matrix can be related to Faddeev components
in the matrix form  by introducing four-dimensional vector
\begin{equation}
\bm{D}=\left(
\begin{array}{c}
1\\1\\1\\1
\end{array}
\right)\, .
\end{equation}
To sum over the channels one has to multiplying the corresponding
matrix or vector by the vector $\bm{D}^T$ and/or $\bm{D}$
\begin{eqnarray}
{\cal T}&=&\sum_{i,j=0}^3T^{ij}=\bm{{D}}^T\bm{{\cal T}}\bm{{D}} , \\
\left|\Psi \right> &=& \bm{{D}}^T \left| \bm{\Psi} \right> \, ,
\quad \quad \left<\Psi \right| = \left< \bm{\Psi} \right| \bm{{D}} , \\
\left|\Gamma \right> &=& \bm{{D}}^T \left| \bm{\Gamma} \right> \, ,
\quad \quad \left<\Gamma \right| = \left< \bm{\Gamma} \right| \bm{{D}} \, .
\end{eqnarray}

Substituting this into the six-point function (\ref{Gresolved})
\begin{equation}
{\cal G}=G^0_{BS}-G^0_{BS}\bm{{D}}^T\bm{{\cal T}}\bm{{D}}G^0_{BS} , ,
\end{equation}
and using
\begin{eqnarray}
\bm{{D}}^T\bm{{D}}&=&4 \, , \\
(\bm{{1+B}})\bm{{D}}&=&4\bm{{D}} \, ,
\label{onePlusBD}
\end{eqnarray}
the six-point propagator (\ref{Gresolved}) is re-written as
\begin{eqnarray}
{\cal G}&=&\frac{1}{4}G^0_{BS}\bm{{D}}^T\bm{{D}}
-G^0_{BS}\bm{{D}}^T\bm{{\cal T}}
\frac{1}{4}\left(\bm{{1+B}}\right)\bm{{D}}G^0_{BS} \nonumber\\
&=&\frac{1}{4}\bm{{D}}^T\left[\bm{{G}}^0_{BS}-\bm{{G}}^0_{BS}
\bm{{\cal T}}\left(\bm{{1+B}}\right)\bm{{G}}^0_{BS}\right]\bm{{D}}
\end{eqnarray}
where $\bm{{G}}^0_{BS}=G^0_{BS}\bm{1}$. Note that in using
(\ref{onePlusBD}) in obtaining this result we could have as easily used
the transpose of this relation which would reverse the order of
${\cal T}$ and $\bm{{1+B}}$.  We can therefore define two matrix propagators
as
\begin{equation}
\bm{{\cal G}}_R=\bm{{G}}^0_{BS}-\bm{{G}}^0_{BS} \bm{{\cal T}}
\left(\bm{{1+B}}\right)\bm{{G}}^0_{BS}
\label{GR_BS}
\end{equation}
and
\begin{equation}
\bm{{\cal G}}_L=\bm{{G}}^0_{BS}-\bm{{G}}^0_{BS}
\left(\bm{{1+B}}\right)\bm{{\cal T}}\bm{{G}}^0_{BS}
\label{GL_BS}\,.
\end{equation}
An additional useful form of the six-point propagator is obtained by
substituting (\ref{t_BS2}) into (\ref{GR_BS})
\begin{equation}
\bm{{\cal G}}_R=\bm{{G}}^0_{BS}-\bm{{G}}^0_{BS} \bm{{\cal
V}}\left(\bm{{1+B}}\right)\bm{{\cal G}}_R
\label{GR_BS2}
\end{equation}
and similarly from (\ref{GL_BS}) one gets
\begin{equation}
\bm{{\cal G}}_L=\bm{{G}}^0_{BS}-\bm{{\cal G}}_L
\left(\bm{{1+B}}\right)\bm{{\cal V}}\bm{{G}}^0_{BS}
\label{GL_BS2}\,.
\end{equation}
The inverses of these propagators are
\begin{eqnarray}
\bm{{\cal G}}^{-1}_R&=&({\bm{{G}}^0_{BS}})^{-1}+ \bm{{\cal V}}
\left(\bm{{1+B}}\right)\, ,
\label{GR_BSinv}\\
\bm{{\cal G}}^{-1}_L&=&({\bm{{G}}^0_{BS}})^{-1}+
\left(\bm{{1+B}}\right)\bm{{\cal V}}
\label{GL_BSinv}\, .
\end{eqnarray}
By construction $\bm{{D}}^T \bm{{\cal G}}_{R,L} \bm{{D}}= 4 {\cal G}$ and
$\bm{{D}}^T \bm{{\cal G}}^{-1}_{R,L} \bm{{D}}= 4 {\cal G}^{-1}$.

The bound and scattering state wave functions can be obtained from
consideration of the singularities of the t-matrix and propagator.
First the three-body t-matrix has simple poles at the location of bound
states.  The matrix form of the t-matrix can then be represented as
\begin{equation}
\bm{{\cal T}}
=\frac{\left|\bm{{\Gamma}}\right>\left<\bm{{\Gamma}}\right|}{P^2-M^2+i\eta}+
\bm{{\cal R}}
\label{3bodyPole}
\end{equation}
where $\left|\bm{{\Gamma}}\right>$ is a vector  of the bound state
vertex functions and $\bm{{\cal R}}$ is a matrix of the residual parts
of the t-matrix. Substituting (\ref{3bodyPole}) into
(\ref{t_BS2}) and extracting the residues of the bound-state
poles yields the equation for the vertex function
\begin{eqnarray}
\left|\bm{{\Gamma}}\right>&=&-\bm{{\cal V}}\bm{{G}}^0_{BS}\left(
\bm{1}+\bm{{B}}\right)\left|\bm{{\Gamma}}\right> \, ,\\
\left<\bm{{\Gamma}}\right|&=&-\left<\bm{{\Gamma}}\right|\left(
\bm{1}+\bm{{B}}\right)\bm{{G}}^0_{BS}\bm{{\cal V}} \, .
\end{eqnarray}
Defining, the Bethe-Salpeter wave function as
$\left|\bm{{\Psi}}\right>= \bm{{G}}^0_{BS}\left|\bm{{\Gamma}}\right>$,
this can be rewritten as
\begin{eqnarray}
\left[ ({\bm{{G}}^0_{BS}})^{-1}+\bm{{\cal V}}\left(
\bm{1}+\bm{{B}}\right)\right]\left|\bm{{\Psi}}\right>&=&
\bm{{\cal G}}^{-1}_R\left|\bm{{\Psi}}\right>=0 \, , \\
\left<\bm{{\Psi}}\right|\left[ ({\bm{{G}}^0_{BS}})^{-1}+\left(
\bm{1}+\bm{{B}}\right)\bm{{\cal V}}\right]&=&
\left<\bm{{\Psi}}\right|\bm{{\cal G}}^{-1}_L=0 \, .
\end{eqnarray}
The normalization condition for the vertex function can be obtained
from the nonlinear form of the t-matrix equation
\begin{equation}
\bm{{\cal T}}=\bm{{\cal V}}-\bm{{\cal T}}\bm{{G}}^0_{BS}\left(
\bm{{1}}+\bm{{B}}\right)\bm{{\cal T}}-\bm{{\cal T}}\left(
\bm{{1}}+\bm{{B}}\right)\bm{{G}}^0_{BS}\bm{{\cal
V}}\bm{{G}}^0_{BS}\left( \bm{{1}}+\bm{{B}}\right)\bm{{\cal T}}\,.
\label{nonlinear_BS}
\end{equation}
Substituting (\ref{3bodyPole}) into (\ref{nonlinear_BS}) and isolating
the residues of the simple pole yields
\begin{eqnarray}
1&=&-\left<\bm{{\Gamma}}\right|\left.\frac{\partial}{\partial
P^2}\left[\bm{{G}}^0_{BS}\left(
\bm{{1}}+\bm{{B}}\right)+\bm{{G}}^0_{BS}\left(
\bm{{1}}+\bm{{B}}\right)\bm{{\cal V}}\left(
\bm{{1}}+\bm{{B}}\right)\bm{{G}}^0_{BS}\right]\right|_{P^2=M^2}
\left|\bm{{\Gamma}}\right>
\nonumber\\
&=&\left<\bm{{\Gamma}}\right|\left[\frac{\partial\bm{{G}}^0_{BS}}{\partial
P^2}\left( \bm{{1}}+\bm{{B}}\right)-\bm{{G}}^0_{BS}\left(
\bm{{1}}+\bm{{B}}\right)\frac{\partial\bm{{\cal V}}}{\partial
P^2}\left(
\bm{{1}}+\bm{{B}}\right)\bm{{G}}^0_{BS}\right]_{P^2=M^2}\left|
\bm{{\Gamma}}\right>\,.
\end{eqnarray}
Since
\begin{equation}
\bm{{D}}\bm{{D}}^T=\bm{{1}}+\bm{{B}}\,,
\end{equation}
this is equivalent to the normalization condition given in ref.\
\cite{norm}.

The outgoing scattering wave function (\ref{scat3BS}) can be re-written
as
\begin{equation}
\left<\Psi^{(-)}\right|=\frac{1}{4}\left<\bm{p}_1\bm{p}_2\bm{p}_3
\right| \bm{{D}}^T \left(\bm{1}-\left(\bm{1}+\bm{B}\right)\bm{{\cal T}}
\bm{G}^0_{BS}\right)\bm{D} \, ,
\end{equation}
from which one gets the corresponding vector in the space of Faddeev
components:
\begin{equation}
\left<\bm{{\Psi}}^{(-)}\right|=\frac{1}{4}\left<\bm{p}_1\bm{p}_2\bm{p}_3
\right| \bm{{D}}^T\left[\bm{{1}}-
\left(\bm{{1}}+\bm{{B}}\right)\bm{{\cal T}}\bm{{G}}^0_{BS}\right] \, ,
\label{3s_out4}
\end{equation}
where
\begin{equation}
\left<\bm{{\Psi}}^{(-)}\right|\bm{{\cal G}}_L^{-1}=0\,.
\end{equation}
Similarly, the incoming scattering wave function ({\ref{scat3BSin}) is
\begin{equation}
\left|\bm{{\Psi}}^{(+)}\right>=\frac{1}{4}\left[\bm{{1}}-
\bm{{G}}^0_{BS}\bm{{\cal
T}}\left(\bm{{1}}+\bm{{B}}\right)\right]\bm{{D}}
\left|\bm{p}_1\bm{p}_2\bm{p}_3\right>
\end{equation}
where
\begin{equation}
\bm{{\cal G}}_R^{-1}\left|\bm{{\Psi}}^{(+)}\right>=0\,.
\end{equation}

The two-body scattering state for the three-body system can be obtained
from ({\ref{scat2BS}) as
\begin{equation}
\left<\Phi^{1(-)}\right|=\left<\Phi^{(2)1},\bm{p}_1\right| \left(
\begin{array}{cccc}
0 &  1 & 0 & 0
\end{array}
\right)\left(\bm{1}-\bm{B}\bm{{\cal T}}\bm{G}^0_{BS}\right)\bm{D}\,.
\end{equation}
The vector form of this state can then be identified as
\begin{equation}
\left<\bm{\Phi}^{1(-)}\right|=\left<\Phi^{(2)1},\bm{p}_1\right|
\left(
\begin{array}{cccc}
0 & 1 & 0 & 0
\end{array}
\right)\left(\bm{1}-\bm{B}\bm{{\cal T}}\bm{G}^0_{BS}\right)\,.
\label{2s_out4}
\end{equation}
where
\begin{equation}
\left<\bm{\Phi}^{1(-)}\right|\bm{{\cal G}}^{-1}_L=0\,.
\end{equation}
The initial two-body scattering wave function is
\begin{equation}
\left|\bm{\Phi}^{1(+)}\right>=\left(\bm{1}-\bm{G}^0_{BS}
\bm{{\cal T}\bm{B}}\right) \left(
\begin{array}{c}
0 \\ 1  \\ 0 \\ 0
\end{array}
\right) \left|\Phi^{(2)1},\bm{p}_1\right>\,.
\end{equation}
where
\begin{equation}
\bm{{\cal G}}^{-1}_R\left|\bm{\Phi}^{1(+)}\right>=0\,.
\end{equation}

%--------------------------------------------------------
\subsection{Seven-point vertex function and current operators}

The matrix form of the seven-point propagator can now be defined such
that
\begin{equation}
{\cal G}^\mu=\bm{{D}}^T\bm{{\cal G}}^\mu\bm{{D}}.
\end{equation}
Using this and the properties of $\bm{{D}}$, the matrix form of the
seven-point function is given by
\begin{eqnarray}
\bm{{\cal G}}^\mu&=&-\frac{1}{16}\Biggl\{\left[\bm{{1}}-
G^0_{BS}\left(\bm{{1+B}}\right) \bm{{\cal T}}
\right]G^0_{BS}J^{(1)\mu}G^0_{BS}\left(\bm{{1+B}}\right)
\left[\bm{{1}}-\bm{{\cal T}}G^0_{BS}\left(\bm{{1+B}}\right)\right]
\nonumber\\ &&+G^0_{BS}\left(\bm{{1+B}}\right)\left[\bm{{1}}-\bm{{\cal
T}} G^0_{BS}\left(\bm{{1+B}}\right)\right]\bm{J}^\mu_{\rm int}
\left[\bm{{1}}-G^0_{BS}\left(\bm{{1+B}}\right)\bm{{\cal T}}\right]
G^0_{BS}\left(\bm{{1+B}}\right)\Biggr\}\nonumber\\
&=&-\frac{1}{16}\bm{{\cal G}}_L
\left[J^{(1)\mu}\left(\bm{{1+B}}\right)+
\left(\bm{{1+B}}\right)\bm{J}^\mu_{\rm int}\left(\bm{{1+B}}\right)
\right]\bm{{\cal G}}_R
\end{eqnarray}
where
\begin{equation}
J^{(1)\mu}=\sum_{i=1}^3 J^{i\mu} \, ,
\end{equation}
and we have introduced the diagonal matrix with components defined by
(\ref{Jtint},\ref{J0int}):
\begin{eqnarray}
\bm{J}^\mu_{\rm int}&=& diag\left( J^{0\mu}_{\rm int}, J^{1\mu}_{\rm int},
J^{2\mu}_{\rm int}, J^{3\mu}_{\rm int} \right) \, , \\
q_\mu \bm{J}^\mu_{\rm int} &=&  \left[ e_T(q), \bm{{\cal V}} \right] \, .
\end{eqnarray}
The effective current can then be identified as
\begin{equation}
\bm{J}^\mu_{\rm eff}=J^{(1)\mu}\left(\bm{{1+B}}\right)+
\left(\bm{{1+B}}\right)\bm{J}^\mu_{\rm int}\left(\bm{{1+B}}\right)
\, .
\label{Jeffmat}
\end{equation}
Since $\bm{J}^\mu_{\rm int}$ is diagonal, the last term can be {\em identically}
re-written as $\left(\bm{{1+B}}\right)J^\mu_{\rm int}$, but the form given above is
more convenient for discussion below.
Contraction of the four-momentum transfer with the effective current
gives
\begin{eqnarray}
q_\mu\bm{J}^\mu_{\rm eff}
&=&\left[e_T(q),{G^0_{BS}}^{-1}\right]\left(\bm{{1+B}}\right)+
\left(\bm{{1+B}}\right)\left[e_T(q),\bm{{\cal V}}\right]
\left(\bm{{1+B}}\right)\nonumber\\
&=&e_T(q)\left(\bm{{1+B}}\right)\bm{{\cal G}}_R^{-1}-\bm{{\cal
G}}_L^{-1}\left(\bm{{1+B}}\right)e_T(q)\,.
\label{WTmat}
\end{eqnarray}
So the current will be conserved.

%___________________________________________________________________

%\renewcommand{\theequation}{B.\arabic{equation}}
%\setcounter{equation}{0}
%\section*{Appendix B: Matrix form of the BS equation: Identical Particles}
%\setcounter{subsection}{0}
\section{Matrix form of the BS equation: Identical Particles}
\subsection{Propagators and wave functions}

To symmetrize the matrix form of the Bethe-Salpeter amplitudes (\ref{t_BS},
\ref{t_BS2}) we consider first the diagonal matrices
\begin{equation}
\bm{{\cal Z}}=
diag (Z^0, Z^1 i G_1^{-1}, Z^2 i G_2^{-1}, Z^3 i G_3^{-1}) \, ,
\label{Zdiag}
\end{equation}
where $Z= V$ or $M$. Let us first symmetrize the components of this matrix
in respect to the exchange of interacting particles defining:
\begin{equation}
\bm{\overline{Z}}= \bm{{\cal A}}_{ip}\, \bm{{\cal Z}}=
 diag({\cal A}_3 Z^0, \overline{Z}^1 i G_1^{-1},
\overline{Z}^2 i G_2^{-1},\overline{Z}^3 i G_3^{-1} )
\, ,
\end{equation}
where (making use of $P_{jk} Z^i P_{jk}= Z^i$)
\begin{equation}
\overline{Z}^i= {\cal A}_2^i Z^i= Z^i {\cal A}_2^i=
{\cal A}_2^i Z^i {\cal A}_2^i
\end{equation}
and the diagonal matrix ${\cal A}_{ip}$ is expressed in terms of the
symmetrization operators (\ref{symmetry3},\ref{A2k}):
\begin{equation}
{\cal A}_{ip}= diag({\cal A}_3, {\cal A}_2^1, {\cal A}_2^2, {\cal A}_2^3)
\, .
\end{equation}
Since
\begin{equation}
\bm{{\cal A}}_{ip} \bm{{\cal A}}_{ip} = \bm{{\cal A}}_{ip}
\, ,
\end{equation}
we can write
\begin{equation}
\overline{\bm{Z}}= \bm{{\cal A}}_{ip} \bm{{\cal Z}} =
\bm{{\cal Z}} \bm{{\cal A}}_{ip} =
\bm{{\cal A}}_{ip}\bm{{\cal Z}}\bm{{\cal A}}_{ip} \, ,
\end{equation}
and get the symmetrized equations for m-matrices in the form:
\begin{equation}
\bm{\overline{M}}= \bm{\overline{V}}-
\bm{\overline{V}} G^0_{BS} \bm{\overline{M}}=
\bm{\overline{V}}- \bm{\overline{M}}G^0_{BS} \bm{\overline{V}} \, .
\label{Mmat}
\end{equation}
Symmetrizing inside each channel, ie.\ in respect to interchange
of interacting particles, does not, of course, account for the
total symmetrization, in particular, eg.\ $\bm{D}^T\,
\bm{\overline{V}}\bm{D} \neq V= {\cal A}_3 {\cal V}$, where $V$
defined in eq.\ (\ref{Vsym}) and enters as the driving term into
equation for symmetrized t-matrix (\ref{tmatsym}). It is necessary
to introduce the matrix $\bm{{\cal A}}_s$ that symmetrizes between
individual channels:
\begin{equation}
\bm{{\cal A}}_s=\left(
\begin{array}{cccc}
1 & 0 & 0 & 0 \\
0 & \frac{1}{3} & \frac{1}{3}\zeta {\cal P}_{12} &
\frac{1}{3}\zeta {\cal P}_{13}
\\
0 & \frac{1}{3}\zeta {\cal P}_{12} & \frac{1}{3} &
\frac{1}{3}\zeta {\cal P}_{23} \\
0 & \frac{1}{3}\zeta {\cal P}_{13} &
\frac{1}{3}\zeta {\cal P}_{23} & \frac{1}{3} \\
\end{array}
\right)
\end{equation}
satisfying
\begin{equation}
\bm{{\cal A}}_s\, \bm{D}= \left(
\begin{array}{c}
1  \\
\Pi_1  \\
\Pi_2  \\
\Pi_3   \\
\end{array}
\right)
\end{equation}
where $\Pi_i$ are defined by eq.\ (\ref{Pik}). It is easy to see that
\begin{equation}
\bm{D}^T\, \bm{{\cal A}}_s\, \bm{\overline{Z}}\bm{D} =
{\cal A}_3 {\cal Z}= {\cal A}_3 Z^0 +
\sum_{i=1}^3 \Pi_i\, \overline{Z}^i\, i G^{-1}_i = Z \, ,
\end{equation}
as required by eq.\ (\ref{Vsym}). The fully symmetrized matrix
$\bm{Z}$ is therefore given by
\begin{eqnarray}
\bm{Z}&=& \bm{{\cal A}}_s\, \bm{\overline{Z}}=
(\bm{{\cal A}}_s \bm{{\cal A}}_{ip})\, \bm{{\cal Z}}=
\bm{{\cal A}} \, \bm{{\cal Z}} \nonumber\\
&=&
\left(
\begin{array}{cccc}
{\cal A}_3 Z^0 & 0 & 0& 0 \\
0 & \frac{1}{3}\overline{Z}^1 iG_1^{-1} &
\frac{1}{3}\zeta {\cal P}_{12}\overline{Z}^2 iG_2^{-1}&
\frac{1}{3}\zeta {\cal P}_{13}\overline{Z}^3 iG_3^{-1} \\
0 & \frac{1}{3}\zeta {\cal P}_{12}\overline{Z}^1 iG_1^{-1}&
\frac{1}{3}\overline{Z}^2 iG_2^{-1} &
\frac{1}{3}\zeta {\cal P}_{23}\overline{Z}^3 iG_3^{-1} \\
0 & \frac{1}{3}\zeta {\cal P}_{13}\overline{Z}^1 iG_1^{-1}&
\frac{1}{3}\zeta {\cal P}_{23}\overline{Z}^2 iG_2^{-1} &
\frac{1}{3}\overline{Z}^3 iG_3^{-1} \\
\end{array}
\right)
\, ,
\label{Zmsym}
\end{eqnarray}
where the full symmetrization matrix reads:
\begin{eqnarray}
\bm{{\cal A}}&=& \bm{{\cal A}}_s \bm{{\cal A}}_{ip}=
\left(
\begin{array}{cccc}
{\cal A}_3& 0   & 0 & 0\\
0&\frac{1}{3}{\cal A}_2^1 & \frac{1}{3}\zeta {\cal P}_{12}{\cal A}_2^2&
\frac{1}{3}\zeta {\cal P}_{13}{\cal A}_2^3 \\
0&\frac{1}{3}\zeta {\cal P}_{12}{\cal A}_2^1& \frac{1}{3}{\cal A}_2^2&
\frac{1}{3}\zeta {\cal P}_{23}{\cal A}_2^3\\
0& \frac{1}{3}\zeta {\cal P}_{13}{\cal A}_2^1&
\frac{1}{3}\zeta {\cal P}_{23}{\cal A}_2^2& \frac{1}{3}{\cal A}_2^3
\end{array}
\right) \nonumber\\
&=&  \bm{{\cal A}}_{ip} \bm{{\cal A}}_s =
\left(
\begin{array}{cccc}
{\cal A}_3& 0   & 0 & 0\\
0&\frac{1}{3}{\cal A}_2^1 & \frac{1}{3}{\cal A}_2^1 \zeta {\cal P}_{12}&
\frac{1}{3}{\cal A}_2^1 \zeta {\cal P}_{13}\\
0&\frac{1}{3}{\cal A}_2^2 \zeta {\cal P}_{12}& \frac{1}{3}{\cal A}_2^2&
\frac{1}{3}{\cal A}_2^2 \zeta {\cal P}_{23}\\
0& \frac{1}{3}{\cal A}_2^3 \zeta {\cal P}_{13}&
\frac{1}{3}{\cal A}_2^3 \zeta {\cal P}_{23}& \frac{1}{3}{\cal A}_2^3
\end{array}
\right) \, .
\label{Amat}
\end{eqnarray}
When summed over channels it reduces to the full symmetrization
operator ${\cal{A}}_3$:
\begin{equation}
\bm{{\cal A}}\bm{D} = \bm{D}{\cal{A}}_3 \, , \quad \quad
\bm{D}^T \bm{{\cal A}} = \bm{D}^T {\cal{A}}_3 \, ,
\label{Achan}
\end{equation}
and with other matrices introduced above it satisfies
the identities:
\begin{eqnarray}
 \bm{{\cal A}}_{ip}\bm{{\cal A}} &=&\bm{{\cal A}}\bm{{\cal A}}_{ip} =
 \bm{{\cal A}} \, ,\\
 \bm{{\cal A}}_s \bm{{\cal A}} &=&\bm{{\cal A}}\bm{{\cal A}}_s =
 \bm{{\cal A}} \, ,\\
 \bm{{\cal A}}^2 &=& \bm{{\cal A}} \, , \\
 \bm{{\cal A}}\bm{B}&=&\bm{B}\bm{{\cal A}} \, .
\end{eqnarray}
Note that the matrices $ \bm{{\cal A}}_s$
and $\bm{{\cal A}}_{ip}$ do not commute with the matrix $\bm{B}$,
although $\bm{{\cal A}}= \bm{{\cal A}}_s \bm{{\cal A}}_{ip}$
does. From these identities several useful relations follow:
\begin{eqnarray}
\bm{{\cal A}}\, \bm{\overline{Z}}&=&
\bm{\overline{Z}}\, \bm{{\cal A}} =
\bm{{\cal A}}\, \bm{\overline{Z}}\, \bm{{\cal A}} \, , \\
\bm{Z}&=&  \bm{{\cal A}}\, \bm{{\cal Z}}=
\bm{{\cal A}}_s\, \bm{\overline{Z}}=
\bm{\overline{Z}}\, \bm{{\cal A}}_s =
\bm{{\cal A}}_s\, \bm{\overline{Z}}\, \bm{{\cal A}}_s =
 \bm{{\cal A}}\, \bm{\overline{Z}}=
\bm{\overline{Z}}\, \bm{{\cal A}} \, ,
\label{Zsym}
\end{eqnarray}
Now we can symmetrize any general (non-diagonal) matrix in 4-channel space,
eg.\ the three particle t-matrix ${\cal T}$:
\begin{eqnarray}
\bm{{T}}&=& \bm{{\cal A}}\bm{{\cal T}}\bm{{\cal A}} \, , \\
T&=& {\cal A}_3 {\cal T}= \bm{D}^T\, \bm{{T}} \bm{D} \, .
\end{eqnarray}
The symmetrized matrix satisfies the identities
\begin{eqnarray}
\bm{T}&=&\bm{{\cal A}}\bm{T}=\bm{T}\bm{{\cal A}}=
\bm{{\cal A}}\bm{T}\bm{{\cal A}}\\
\bm{T}&=&\bm{{\cal A}}_{ip}\bm{T}=\bm{T}\bm{{\cal A}}_{ip}
=\bm{{\cal A}}_{ip}\bm{T}\bm{{\cal A}}_{ip}\\
\bm{T}&=&\bm{{\cal A}}_s\bm{T}=\bm{T}\bm{{\cal A}}_s=
\bm{{\cal A}}_s\bm{T}\bm{{\cal A}}_s\, ,
\end{eqnarray}
which allows to cast the matrix equation for the symmetrized t-matrix
into the form:
\begin{eqnarray}
\bm{T}&=& \bm{M}- \bm{M} G^0_{BS} \bm{B} \bm{T}=
\bm{M}- \bm{T}\bm{B}G^0_{BS} \bm{M}
\label{TmatM0} \\
  &=& \bm{V} - \bm{V} G^0_{BS}(\bm{{1+B}}) \bm{T}=
 \bm{V} -\bm{T}(\bm{{1+B}})G^0_{BS} \bm{V}\, .
\label{TmatM1}
\end{eqnarray}
Since according to (\ref{Zsym}) $\bm{Z}= \bm{{\cal A}}_s\, \bm{\overline{Z}}=
\bm{{\cal A}}\, \bm{\overline{Z}}$ and matrices $\bm{A}$ and $\bm{B}$ commute,
one can re-write these equations into more convenient equivalent form
in terms of diagonal matrices $\bm{\overline{M}}$ and $\bm{\overline{V}}$:
\begin{eqnarray}
\bm{T}&=& \bm{\overline{M}}\bm{{\cal A}}_s -
\bm{\overline{M}} G^0_{BS} \bm{B} \bm{T}=
\bm{\overline{M}}\bm{{\cal A}}_s -
 \bm{T}\bm{B}G^0_{BS} \bm{\overline{M}}
\label{TmatM} \\
   &=& \bm{\overline{V}}\bm{{\cal A}}_s -
\bm{\overline{V}} G^0_{BS}(\bm{{1+B}}) \bm{T}=
 \bm{\overline{V}}\bm{{\cal A}}_s
-\bm{T}(\bm{{1+B}})G^0_{BS} \bm{\overline{V}}\, , \\
 &=& \bm{\overline{V}}\bm{{\cal A}}_s -
 \bm{T} G^0_{BS} (\bm{{1+B}}) \bm{T} -
 \bm{T} (\bm{{1+B}}) G^0_{BS} \bm{\overline{V}} G^0_{BS}
(\bm{{1+B}}) \bm{T} \, .
\end{eqnarray}

Proceeding exactly the same way as in the previous appendix one
gets from these relations the dynamical equations and
normalization conditions for the symmetrized bound state amplitudes:
\begin{eqnarray}
\left|\bm{{\Gamma}}_s\right> &=& \bm{{\cal A}}
\left|\bm{{\Gamma}}\right> \, , \\
\left|\bm{{\Gamma}}_s\right>&=& \bm{{\cal A}}\left|\bm{{\Gamma}}_s\right>
=\bm{{\cal A}}_s\left|\bm{{\Gamma}}_s\right>=
\bm{{\cal A}}_{ip}\left|\bm{{\Gamma}}_s\right> \, , \\
\left|\bm{{\Gamma}}_s\right>&=& -\bm{V} G^0_{BS}
\left(\bm{1}+\bm{{B}}\right)\left|\bm{{\Gamma}}_s\right>=
- \bm{\overline{V}}  G^0_{BS}\left(
\bm{1}+\bm{{B}}\right)\left|\bm{{\Gamma}}_s\right> \, ,
\label{Gamma4v}\\
1&=& \left<\bm{{\Gamma}}_s\right|\left[
\frac{\partial G^0_{BS}}{\partial P^2} \left( \bm{{1}}+\bm{{B}}\right)
- G^0_{BS} \left(\bm{{1}}+\bm{{B}}\right)
\frac{\partial\overline{\bm{V}}}{\partial P^2}
\left(\bm{{1}}+\bm{{B}}\right)  G^0_{BS}
\right]_{P^2=M^2}
\left|\bm{{\Gamma}}_s\right> \, .
\label{Gamma4norm}
\end{eqnarray}

The symmetrized bound state wave function is defined by
\begin{equation}
\left|\bm{{\Psi}}_{Bs}\right>= {G}^0_{BS} \left|\bm{{\Gamma}}_s\right> \, .
\end{equation}
The scattering outgoing states are obtained in a straightforward way acting
by the symmetrization matrix $\bm{{\cal A}}$ on vectors
(\ref{3s_out4},\ref{2s_out4}) and using the fact that
$\bm{{\cal A}}$ commutes with $(\bm{1}+\bm{B})$:
\begin{eqnarray}
\left<\bm{{\Psi}}^{(-)}_s\right| &=& \left<\bm{{\Psi}}^{(-)}\right|
\bm{{\cal A}}=
\frac{1}{4} \left<\bm{p}_1,s_1;\bm{p}_2,s_2;\bm{p}_3,s_3\right|
\bm{{D}}^T\bm{{\cal A}}\left[\bm{{1}}-\left(\bm{{1}}+\bm{{B}}\right)
\bm{ T} G^0_{BS}\right] \, , \\
\left<\bm{\Phi}^{1(-)}_s\right| &=& \left<\bm{\Phi}^{1(-)}\right|
\bm{{\cal A}}= \left<\Phi^{(2)1};\bm{p}_1,s_1\right| \left(
\begin{array}{cccc}
0 & 1 & 0 & 0
\end{array}
\right)
\bm{{\cal A}} \left(\bm{1}-\bm{B}\bm{T} G^0_{BS} \right) \, .
\end{eqnarray}
From eqs.\ (\ref{GR_BS},\ref{GL_BS}) one gets the symmetrized
propagators
\begin{eqnarray}
\bm{G}_R&=& G^0_{BS}\bm{{\cal A}}
- G^0_{BS}\bm{T}(\bm{{1+B}}) G^0_{BS}\bm{{\cal A}} \, ,
\label{GRsym} \\
\bm{G}_L&=& G^0_{BS}\bm{{\cal A}}- G^0_{BS}\bm{{\cal A}}
(\bm{{1+B}})\bm{T} G^0_{BS} \, ,
\label{GLsym}
\end{eqnarray}
and from eqs.\ (\ref{GR_BSinv},\ref{GL_BSinv}) the inverse
propagators
\begin{eqnarray}
\bm{G}_R^{-1}&=& \bm{{\cal A}}\bm{{\cal G}}_R^{-1}=
(G^0_{BS})^{-1}\bm{{\cal A}}+ \bm{V} (\bm{1}+ \bm{B})
 \, ,
\label{GRsyminv} \\
\bm{G}_L^{-1}&=& \bm{{\cal A}}\bm{{\cal G}}_L^{-1}=
(G^0_{BS})^{-1}\bm{{\cal A}}+ (\bm{1}+ \bm{B})\bm{V}
 \, ,
\label{GLsyminv}
\end{eqnarray}
which satisfy
\begin{eqnarray}
\bm{G}_R^{-1} \bm{G}_R &=& \bm{G}_R \bm{G}_R^{-1}= \bm{{\cal A}} \, , \\
\bm{G}_L^{-1} \bm{G}_L &=& \bm{G}_L \bm{G}_L^{-1}= \bm{{\cal A}} \, ,
\end{eqnarray}
and in terms of which the generic equations for the symmetrized
wave functions read
\begin{eqnarray}
\bm{G}_R^{-1} \left|\bm{{\Psi}}_s\right>  &=&  0 \, , \\
\left< \bm{{\Psi}}_s\right| \bm{G}_L^{-1}  &=&  0 \, .
\end{eqnarray}
%

%-----------------------------------------------------------
%-----------------------------------------------------------
%-----------------------------------------------------------

Finally the 7-point function is:
\begin{eqnarray}
\bm{{G}}^\mu&=&\bm{{\cal A}}\bm{{\cal G}}^\mu\nonumber\\
&=&-\frac{1}{16}\bm{G}_L \left[J^{(1)\mu}\left(\bm{{1+B}}\right)+
\left(\bm{{1+B}}\right)\bm{J}^\mu_{\rm int}\left(\bm{{1+B}}\right)
\right]\bm{G}_R
\label{gmusym}
\end{eqnarray}
%

%-----------------------------------------------------------
\subsection{Reducing to two independent channels}

The matrix elements of the symmetrized operators and components
of the symmetrized wave/vertex functions are connected by symmetry
relations:
\begin{eqnarray}
S^{0m}=\zeta{\cal P}_{ij} S^{0m} \quad \quad S^{m0}=
 S^{m0} \zeta{\cal P}_{ij} \, & , &
\quad
\left| \Psi_s^0 \right>=\zeta{\cal P}_{ij} \left| \Psi_s^0\right> \, ,
\quad
\left| \Gamma_s^0\right>=\zeta{\cal P}_{ij} \left| \Gamma_s^0\right> \, ,
\label{Sym0}\\
S^{ir}=\zeta{\cal P}_{jk} S^{ir}= S^{ir} \zeta{\cal P}_{st} \, & , &
\quad
\left| \Psi_s^i \right>=\zeta{\cal P}_{jk} \left| \Psi_s^i\right> \, ,
\quad
\left| \Gamma_s^i\right>=\zeta{\cal P}_{jk} \left| \Gamma_s^i\right> \, ,
\label{Symip}\\
S^{ir}=\zeta{\cal P}_{ik} S^{kr}= S^{it} \zeta{\cal P}_{rt} \, & , &
\quad
\left| \Psi_s^i \right>=\zeta{\cal P}_{ik} \left| \Psi_s^k\right> \, ,
\quad
\left| \Gamma_s^i\right>=\zeta{\cal P}_{ik} \left| \Gamma_s^k\right> \, ,
\label{Syms}
\end{eqnarray}
for any $m= 0, 1, 2, 3$ and $i \neq j \neq k \neq 0$ ($r \neq s
\neq t \neq 0$). This means that there are only four independent
components of the symmetrized matrices:
$S^{00},S^{01},S^{10},S^{11}$ and only two independent components
of the symmetrized wave (vertex) functions: $\Psi_s^0, \Psi_s^1$
($\Gamma_s^0, \Gamma_s^1$). The symmetry relations were discussed
in Section III of ref.\ (\cite{norm}) (for bound state wave
function in component form). Let us briefly show how they follow
from matrix formalism used in this paper.

Let us introduce auxiliary matrices:
\begin{eqnarray}
\bm{{\cal A}}_{ips}&=&
diag(\zeta {\cal P}_{ij}, \zeta {\cal P}_{23},
\zeta {\cal P}_{13}, \zeta {\cal P}_{12}) \, , \\
\bm{{\cal A}}_{ss}&=&\left(
\begin{array}{cccc}
1 & 0 & 0 & 0 \\
0 & c_{11} &  c_{12}\zeta {\cal P}_{12} &
 c_{13}\zeta {\cal P}_{13}
\\
0 &  c_{21}\zeta {\cal P}_{12} & c_{22}  &
 c_{23}\zeta {\cal P}_{23} \\
0 &  c_{31}\zeta {\cal P}_{13} &
 c_{32}\zeta {\cal P}_{23} &  c_{33} \\
\end{array}
\right) \, .
\end{eqnarray}
Obviously $\bm{{\cal A}}_{ips} \bm{{\cal A}}_{ip}= \bm{{\cal A}}_{ps}$
for any $i,j$, therefore it holds
$\bm{{\cal A}}_{ips} \bm{{\cal A}}= \bm{{\cal A}}$
and thus:
\begin{eqnarray}
\bm{{\cal A}}_{ips} \left| \bm{\Psi}_s \right>&=&
\left| \bm{\Psi}_s \right> \, , \\
\bm{{\cal A}}_{ips} \bm{S}&=& \bm{S}\bm{{\cal A}}_{ips} = \bm{S} \, ,
\end{eqnarray}
from which the relations (\ref{Sym0},\ref{Symip}) follow immediately.
The matrix $\bm{{\cal A}}_{ss}$ mixes the channels 1,2,3 arbitrarily.
With the help of relations
${\cal P}_{ik}{\cal P}_{ij}={\cal P}_{ij}{\cal P}_{jk}$
and ${\cal A}_2^i \zeta {\cal P}_{jk}= \zeta {\cal P}_{jk}{\cal A}_2^i =
{\cal A}_2^i$ it is easy to show that
\begin{equation}
\bm{{\cal A}}_{ss}\bm{{\cal A}}=
\bm{{\cal A}}_{ss}\bm{{\cal A}}_{s}\bm{{\cal A}}_{ip}= \bm{{\cal A}}=
\bm{{\cal A}} \bm{{\cal A}}_{ss} \, ,
\end{equation}
provided the coefficients $c_{ij}$ satisfy $\sum_i c_{ij}= \sum_i c_{ji}=1$
for any $j$. For the matrices $\bm{{\cal A}}_{ss}$ with so restricted
coefficients $c_{ij}$ it then follows
\begin{eqnarray}
\bm{{\cal A}}_{ss} \left| \bm{\Psi}_s \right>&=&
\left| \bm{\Psi}_s \right> \, , \\
\bm{{\cal A}}_{ss} \bm{S}&=& \bm{S}\bm{{\cal A}}_{ss} = \bm{S} \, ,
\end{eqnarray}
Putting now $c_{12}=c_{23}=c_{31}=1$ and all others equal to zero,
one gets the symmetry relations (\ref{Syms}).

To reduce our 4 channel formalism from the previous section to
2 independent channels, taking into account the symmetry relations
(\ref{Sym0}-\ref{Syms}) it is useful to introduce the matrix
\begin{equation}
\bm{X}=\left(
\begin{array}{cc}
1 & 0\\ 0 & 1\\ 0 & \zeta {\cal P}_{12}\\ 0 & \zeta {\cal P}_{13}
\end{array}
\right)\, ,
\end{equation}
with the help of which one can write
\begin{eqnarray}
\left| \bm{\Psi}_s \right>&=&  \bm{X} \left| \bm{\psi} \right> \, , \\
\left| \bm{\Gamma}_s \right>&=&  \bm{X} \left| \bm{\gamma} \right> \, , \\
\bm{S}&=&\bm{X}\bm{s}\bm{X}^T
\, ,
\label{Ssrel}
\end{eqnarray}
where
\begin{eqnarray}
\left| \bm{\psi} \right>=
\left(
\begin{array}{c}
\left|\Psi_s^0\right> \\ \left|\Psi_s^1\right>
\end{array}
\right)
&,&
\left| \bm{\gamma} \right>=
\left(
\begin{array}{c}
\left|\Gamma_s^0\right> \\ \left|\Gamma_s^1\right>
\end{array}
\right) \, ,
\label{psigam2}\\
\bm{s}&=&\left(
\begin{array}{cc}
S^{00} & S^{01}\\ S^{10} & S^{11}
\end{array}
\right)\,.
\label{ssym2}
\end{eqnarray}
The relation (\ref{ssym2}) holds for all fully symmetrized matrices,
e.g., for $S= T, M, V, \dots$. Note, that in particular for
operators ${\cal Z}$ (diagonal for distinguishable particles,
 e.g., $Z= V, M$)  it follows from eq.\ (\ref{Zmsym}):
\begin{eqnarray}
 {\bm{Z}}&=& \bm{{\cal A}} {\cal Z}=
 \bm{{\cal A}}_s \overline{\bm{Z}}=\bm{X} {\bm{z}}\bm{X}^T \, ,
\label{zs} \\
{\bm{z}}&=& \left(
\begin{array}{cc}
{\cal A}_3 Z^{0} & 0\\
0 & \frac{1}{3} \overline{Z}^{1} i G_1^{-1}
\end{array}
\right)\, ,
\end{eqnarray}

To invert the relation (\ref{Ssrel}) between $\bm{S}$ and $\bm{s}$
we introduce the  matrices
\begin{eqnarray}
\bm{Y}&=&\left(
\begin{array}{cccc}
1 & 0 & 0 & 0\\ 0 & \frac{1}{3} & 0 & 0\\ 0 & 0 & \frac{1}{3}  & 0\\ 0
& 0 & 0 & \frac{1}{3}
\end{array}
\right) \, ,
\label{Ymat} \\
\bm{\alpha}&=&\bm{X}^T\bm{X}=\left(
\begin{array}{cc}
1 & 0\\ 0 & 3
\end{array}
\right)\, ,
\label{Alfmat}
\end{eqnarray}
where $\bm{X}$ and $\bm{Y}$ satisfy
\begin{equation}
\bm{X}^T \bm{Y} \bm{X}=\bm{1} \, .
\label{xyx}
\end{equation}
For the symmetrization matrix we get from (\ref{Amat})
\begin{equation}
\bm{a}_{ip}= \bm{X}^T \bm{Y}\, \bm{{\cal A}} \, \bm{Y}\bm{X}=
\left(
\begin{array}{cc}
{\cal A}_3 & 0\\ 0 & \frac{1}{3} {\cal A}_2^1
\end{array}
\right)=
\left(
\begin{array}{cc}
{\cal A}_3 & 0\\ 0 & \frac{1}{6}\left(1+\zeta P_{23}\right)
\end{array}
\right)
\, .
\label{a2dim}
\end{equation}

To sum over two channels we introduce the 2-dimensional vector
$\bm{D}_2$:
\begin{equation}
  \bm{D}_2 =  \bm{X}^T \bm{D}=
\left(
\begin{array}{c}
1 \\ 3\Pi_1
\end{array}
\right)=
\left(
\begin{array}{c}
1 \\ 1+ \zeta {\cal P}_{12}+ \zeta {\cal P}_{13}
\end{array}
 \right) \, .
\label{dvec}
\end{equation}
The total symmetrized wave function can now be written as
\begin{equation}
\left|  \Psi_s \right>= \bm{D}_2^T \left| \bm{\psi} \right>=
\left|  \Psi_s^0 \right>+
(1+ \zeta {\cal P}_{12}+ \zeta {\cal P}_{13}) \left|\Psi_s^1 \right>
\, ,
\end{equation}
and analogously for the vertex function. Similarly, the full
t-matrix is obtained by
\begin{equation}
T= \bm{D}^T \bm{T} \bm{D}= \bm{D}^T \bm{X}\bm{t}\bm{X}^T \bm{D}=
\bm{D}_2^T \bm{t} \bm{D}_2 \, ,
\end{equation}
and analogously for any fully symmetrized matrix. Notice that
\begin{equation}
 \bm{X} \bm{X}^T \bm{Y} \bm{{\cal A}}=
 \bm{{\cal A}} \bm{Y} \bm{X} \bm{X}^T = \bm{{\cal A}}  \, .
\label{xxy}
\end{equation}
Using this relation we easily get
\begin{eqnarray}
\bm{s}= \bm{X}^T \bm{Y}\, \bm{S} \, \bm{Y}\bm{X} \quad &,&
\quad \bm{S}= \bm{X} \bm{s} \bm{X}^T \, ,
\label{s2dim}\\
\bm{r}\, \bm{\alpha}\, \bm{s}= \bm{X}^T \bm{Y}\, \bm{R} \,
\bm{S}\, \bm{Y}\bm{X} \quad &,& \quad
\bm{R} \,\bm{S}=  \bm{X} \bm{r}\, \bm{\alpha}\,\bm{s} \bm{X}^T
\, .
\end{eqnarray}
Last equation shows that in translating the successive action of
two 4-dimensional matrices $\bm{R}$ and  $\bm{S}$ into 2-channel
formalism one has to include the matrix $\bm{\alpha}$ which
simply takes into account the existence of the three channels
connected by the symmetry relations (\ref{Syms}). Notice in particular
that the relation $\bm{{\cal A}}= \bm{{\cal A}}\, \bm{{\cal A}}$
translates into
\begin{equation}
\bm{a}_{ip} = \bm{a}_{ip}\, \bm{\alpha}\,  \bm{a}_{ip} \, .
\end{equation}

The equation for the vertex function and t-matrix contain also
the matrices $\bm{B}$ and $\bm{1}+ \bm{B}$. Although these matrices are
not symmetrized (their components do not satisfy (\ref{Sym0}-\ref{Syms}),
they are always multiplied by some fully symmetrized matrix ($\bm{V}, \bm{M}$
or $\bm{V}$) from which one can pull out the symmetrization matrix $\bm{{\cal A}}$.
In the transition to the two channel formalism we can therefore consider
instead the matrices $\bm{{\cal A}} \bm{B}$ and $\bm{{\cal A}}(\bm{1}+ \bm{B})$.
Since $\bm{1}+\bm{B}= \bm{D}\bm{D}^T$ and $\bm{{\cal A}}\bm{D}= {\cal A}_3 \bm{D}$,
one gets after simple algebra:
\begin{eqnarray}
\bm{X}^T \bm{Y}\, (\bm{1} + \bm{B}) \, \bm{Y}\bm{X}&=& {\cal A}_3
\left(
\begin{array}{cc}
1 & 1 \\ 1 & 1
\end{array}
\right)
= \bm{a}_{ip}\, \bm{\delta}\, \bm{a}_{ip} \, ,
\label{oneB}\\
\bm{\delta}&=&
\left(
\begin{array}{cc}
1 & 3 \\ 3 & 3(1+ 2\zeta{\cal P}_{12} )
\end{array}
\right) \, .
\label{delta}
\end{eqnarray}
From eqs.\ (\ref{a2dim}) and (\ref{oneB}) it follows
\begin{eqnarray}
\bm{X}^T \bm{Y}\,  \bm{B} \, \bm{Y}\bm{X}&=&
\bm{a}_{ip}\, \bm{\beta}\, \bm{a}_{ip} \, ,
\label{B2dim}\\
\bm{\beta}&=& \bm{\delta}-\bm{\alpha}  =
\left(
\begin{array}{cc}
0 & 3\\ 3 & 6\zeta {\cal P}_{12}
\end{array}
\right)\, .
\label{beta}
\end{eqnarray}
From these relations we obtain for products of the fully symmetrized
matrices $\bm{R}$ or $\bm{S}$ and matrices $\bm{B}$ and $\bm{1}+
\bm{B}$ the following translation  table between 4- and 2-channel
formalism
\begin{eqnarray}
\rm{4 channel} \leftrightarrow \rm{4 channel} \quad && \quad
\rm{4 channel} \leftrightarrow \rm{4 channel} \nonumber\\
(\bm{1}+ \bm{B}) \bm{S} \leftrightarrow \bm{a}_{ip} \, \bm{\delta}\, \bm{s}
\quad && \quad
\bm{B} \bm{S} \leftrightarrow \bm{a}_{ip} \, \bm{\beta}\, \bm{s}
\label{bmult1}\\
\bm{R} (\bm{1}+ \bm{B})  \leftrightarrow \bm{r}\, \bm{\delta}\, \bm{a}_{ip}
\quad && \quad
\bm{R} \bm{B} \leftrightarrow \bm{r}\, \bm{\beta}\, \bm{a}_{ip}   \\
\bm{R} (\bm{1}+ \bm{B}) \bm{S} \leftrightarrow \bm{r}\, \bm{\delta}\, \bm{s}
\quad && \quad
\bm{R} \bm{B} \bm{S} \leftrightarrow \bm{r}\, \bm{\beta}\, \bm{s}
\label{bmult4}
\end{eqnarray}
%

%-----------------------------------------------------------
\subsection{T-matrix and wave functions in two channels}

Making use of relations from the previous subsection we can
now easily re-write the equations for t-matrix and wave functions
in the formalism with two independent channels.

Multiplying the equation (\ref{Mmat}) for the m-matrix with
$\bm{{\cal A}}_s$ and using (\ref{zs}) gives
\begin{equation}
 \bm{m} = \bm{v}- \bm{v}\, G^0_{BS}\, \bm{\alpha}\, \bm{m}=
 \bm{v}- \bm{m}\, G^0_{BS}\, \bm{\alpha}\, \bm{v} \, .
\label{mbar}
\end{equation}
For the t-matrix we immediately get (from (\ref{TmatM0}-\ref{TmatM1})
with the help of relations (\ref{bmult1}-\ref{bmult4}):
\begin{eqnarray}
 \bm{t} &=& \bm{m}- \bm{m}\, G^0_{BS}\, \bm{\beta}\, \bm{t}=
            \bm{m}- \bm{t} G^0_{BS} \bm{\beta} \bm{m}
\label{tmatm2}            \\
        &=&  \bm{v}- \bm{v}\, G^0_{BS}\, \bm{\delta} \bm{t}=
            \bm{v}- \bm{t} G^0_{BS} \bm{\delta} \bm{v}
\label{tmatv2} \\
     &=& \bm{v}- \bm{t} G^0_{BS} \bm{\delta} \bm{t}-
\bm{t} \bm{\delta} G^0_{BS} \bm{v} G^0_{BS} \bm{\delta} \bm{t}
\, ,
\label{tmatnl}
\end{eqnarray}
In a similar way we obtain from eq.\ (\ref{GRsym},\ref{GLsym})
the 2-channel propagators:
\begin{eqnarray}
 \tilde{\bm{g}}_R &=& \bm{X}^T \bm{Y}\,  \bm{G}_R \, \bm{Y}\bm{X}=
 \bm{g}_R\, \bm{a}_{ip}=
\left( G^0_{BS} \bm{1}
-G^0_{BS}\, \bm{t}\, \bm{\delta}\, G^0_{BS} \right) \bm{a}_{ip}\, ,
\label{GR2sym}\\
 \tilde{\bm{g}}_L &=&  \bm{X}^T \bm{Y}\,  \bm{G}_L \, \bm{Y}\bm{X}=
 \bm{g}_L\, \bm{a}_{ip}=
\left( G^0_{BS}
-G^0_{BS}\, \bm{\delta}\,\bm{t}\,  G^0_{BS} \right) \bm{a}_{ip}\, .
\label{GL2sym}
\end{eqnarray}
It is convenient to separate the symmetrization matrix $\bm{a}_{ip}$
from the full propagators $\tilde{\bm{g}}_{R,L}$ and define $\bm{g}_{R,L}$
as above. With the help of (\ref{tmatv2})   one gets for $\bm{g}_{R,L}$
\begin{eqnarray}
 \bm{g}_R &=& G^0_{BS} \bm{1}
-G^0_{BS}\, \bm{t}\, \bm{\delta}\, G^0_{BS} =
G^0_{BS} \bm{1}
-G^0_{BS}\, \bm{v}\, \bm{\delta}\,  \bm{g}_R \, , \\
 \bm{g}_L &=& G^0_{BS} \bm{1}
-G^0_{BS}\, \bm{\delta}\, \bm{t}\,  G^0_{BS} =
G^0_{BS} \bm{1}
- \bm{g}_L \, \bm{\delta}\, \bm{v}\,  G^0_{BS}\,   ,
\end{eqnarray}
and for the inverse matrices
\begin{eqnarray}
\bm{g}_R^{-1}&=& \bm{1}\, ({G^0_{BS}})^{-1} + \bm{v}\, \bm{\delta}\, , \\
\bm{g}_L^{-1}&=& \bm{1}\, ({G^0_{BS}})^{-1} + \bm{\delta}\, \bm{v}\,  .
\end{eqnarray}

For the bound state vertex function we get form (\ref{Gamma4v}):
\begin{equation}
\left|\bm{\gamma}\right>=- \bm{v}\, \bm{\delta}\, G^0_{BS}\, \left|\bm{\gamma}\right>
=- \bm{m}\, \bm{\beta}\, G^0_{BS}\, \left|\bm{\gamma}\right> \, ,
\end{equation}
or equivalently
\begin{equation}
\bm{g}_R^{-1} G^0_{BS} \left|\bm{\gamma}\right>=
\bm{g}_R^{-1}\left|\bm{\psi}\right>= 0 \, .
\end{equation}
The corresponding normalization can be obtained either from
(\ref{Gamma4v}) or from non-linear form of the t-matrix equation
(\ref{tmatnl}):
\begin{equation}
1=\left<\bm{\gamma}\right|\left[\frac{\partial G^0_{BS}}{\partial
P^2}\bm{\delta}-G^0_{BS}\bm{\delta}\frac{\partial \bm{v}}{\partial
P^2}\bm{\delta}G^0_{BS}\right]_{P^2=M^2}\left|\bm{\gamma}\right>\, .
\end{equation}
The 2-channel three particle scattering function is defined
by
\begin{eqnarray}
\left<\bm{{\psi}}^{(-)}\right|&=&
\left<\bm{{\Psi}}^{(-)}_s\right|\, \bm{Y} \bm{X} \nonumber\\
&=& \frac{1}{4}\left<\bm{p}_1,s_1;\bm{p}_2,s_2;\bm{p}_3,s_3\right|\,
(\bm{D}^T \bm{{\cal A}}\bm{Y} \bm{X})\left[
\bm{1}- \bm{\alpha}\, \bm{a}_{ip}\, \bm{\delta}\, \bm{t}\, G^0_{BS}\,
\right]
\nonumber\\
&=& \frac{1}{4}\left<\bm{p}_1,s_1;\bm{p}_2,s_2;\bm{p}_3,s_3\right|\,
{\cal A}_3 \bm{d}^T
\left[\bm{1}- \bm{\delta}\, \bm{t}\, G^0_{BS}\, \right]
\nonumber\\
&=& \left<\bm{p}_1,s_1;\bm{p}_2,s_2;\bm{p}_3,s_3\right|\,
{\cal A}_3\,
\left[\frac{1}{4}\bm{d}^T - \bm{D}_2^T\, \bm{t}\, G^0_{BS}\, \right] \, ,
\label{psi3sc2}
\end{eqnarray}
where
\begin{equation}
\bm{d}^T= \left(
\begin{array}{cc}
1 & 1
\end{array}
\right)
\end{equation}
and we have used
\begin{eqnarray}
\bm{D}^T \bm{{\cal A}}\bm{Y} \bm{X} &=& \bm{D}^T {\cal A}_3\, \bm{Y} \bm{X}=
{\cal A}_3\, \bm{d}^T \, , \\
{\cal A}_3\, \bm{\alpha}\, \bm{a}_{ip} &=& {\cal A}_3\, \bm{1} \, ,\\
{\cal A}_3\, \bm{d}^T \, \bm{\delta} &=& 4 {\cal A}_3\, \bm{D}_2^T \, .
\end{eqnarray}
The scattering wave function satisfies
\begin{equation}
\left<\bm{{\psi}}^{(-)}\right|\, \bm{g}_L^{-1}=0 \, .
\end{equation}
In a similar way:
\begin{eqnarray}
\left<\bm{{\phi}}^{1(-)}\right|&=&
\left<\bm{{\Phi}}^{1(-)}_s\right|\, \bm{Y} \bm{X} \nonumber\\
&=& \left<\Phi^{(2)1},\bm{p}_1\right|\, \left(
\begin{array}{cccc}
0 & 1 & 0 & 0
\end{array}
\right)
\bm{{\cal A}}\bm{Y} \bm{X}\left[
\bm{1}- \bm{\alpha}\, \bm{a}_{ip}\, \bm{\beta}\, \bm{t}\, G^0_{BS}\,
\right]
\nonumber\\
&=& \left<\Phi^{(2)1};\bm{p}_1,s_1\right| \left(
\begin{array}{cc}
0 & 1
\end{array}
\right)\bm{a}_{ip}
\left(\bm{1}-\bm{\beta}\bm{t}\bm{G}^0_{BS}\right)\, ,
\label{phim2d}
\end{eqnarray}
where we have used
\begin{eqnarray}
 \bm{{\cal A}}\bm{Y} \bm{X}&=&\bm{X}\bm{a}_{ip} \, , \\
\left(
\begin{array}{cccc}
0 & 1 & 0 & 0
\end{array}
\right) \bm{X} &=& \left(
\begin{array}{cc}
0 & 1
\end{array}
\right) \, .
\end{eqnarray}
Also this wave function satisfies:
\begin{equation}
\left<\bm{\phi}^{1(-)}\right|\, \bm{g}_L^{-1}=0 \, .
\end{equation}

%\newpage

%-----------------------------------------------------------
\subsection{The BS electromagnetic current for identical particles}

The symmetrized seven point Green's function (\ref{gmusym}) is in two-channel
notation reduced to:
\begin{equation}
\bm{{g}}^\mu=-\frac{1}{16}\bm{a}_{ip}\bm{g}_L \left[
J^{(1)\mu}\bm{\delta}+ \bm{\delta}\bm{j}^\mu_{\rm int}\bm{\delta}
\right]\bm{g}_R\bm{a}_{ip} \, ,
\label{7point2d}
\end{equation}
where
\begin{equation}
\bm{j}^\mu_{\rm eff}=J^{(1)\mu}\bm{\delta}+ \bm{\delta}
\bm{j}^\mu_{\rm int}\bm{\delta}\label{jbs2d}
\end{equation}
with
\begin{equation}
\bm{j}^\mu_{\rm int}= \left(
\begin{array}{cc}
{\cal A}_3 J^{0\mu}_{\rm int} & 0\\
0 & \frac{1}{3}  {\cal A}_2^1 J^{1\mu}_{\rm int}
\end{array}
\right)=
\left(
\begin{array}{cc}
\overline{J}^{0\mu}_{\rm int} & 0\\
0 & \frac{1}{3}  \overline{J}^{1\mu}_{\rm int}
\end{array}
\right)
\, .
\label{jbsint2d}
\end{equation}
From divergences of the one-particle and interaction currents
\begin{eqnarray}
q_\mu J^{(1)\mu}&=&\left[ e_T(q),{G^0_{BS}}^{-1}\right] \, , \\
q_\mu\bm{\delta}\bm{j}^\mu_{\rm int}\bm{\delta}&=& \left[
e_T(q),\bm{\delta}\overline{\bm{v}}\bm{\delta}\right] \, ,
\end{eqnarray}
the two-channel form of the total WT identity follows:
\begin{equation}
q_\mu \bm{j}^\mu_{\rm eff}=e_T(q)\bm{\delta}\bm{g}_R^{-1}
-\bm{g}_L^{-1}e_T(q)\bm{\delta} \, .
\end{equation}

%___________________________________________________________________

\section{Rules for Resolving Products of Operators}

This appendix contains a summary of the rules for resolving
products of singular operators obtained in \cite{2Ncur}.

\begin{description}
\item[Rule 1:]
\begin{equation}
i{\cal Q}_i\,G^{-1}_i\,i{\cal Q}_i\rightarrow\, i{\cal Q}_i\,  .
\label{rule1}
\end{equation}
\item[Rule 2:]
\begin{equation}
i{\cal Q}_i\,G^{-1}_i\,\Delta G_i= \Delta G_i \,G^{-1}_i\, i{\cal
Q}_i \rightarrow 0 \, ,\label{rule2}
\end{equation}
\item[Rule 3:]
\begin{equation}
\Delta G_iG^{-1}_i\Delta G_i \rightarrow \Delta G_i \, .
\label{rule3}
\end{equation}
\end{description}
Note that these identities all refer to products where $G^{-1}_i$
is inserted between factors of ${\cal Z}^1_i=i{\cal Q}_i$ or
${\cal Z}^2_i=\Delta G_i$, and can be summarized by the compact
statement
\begin{eqnarray}
{\cal Z}^\ell_i\,G^{-1}_i\,{\cal Z}^{\ell'}_i\to
\delta_{\ell\ell'}\, {\cal Z}^\ell_i\, . \nonumber
\end{eqnarray}

For operators ${\cal O}$ other than ${\cal Z}^\ell_i$,
\begin{description}
\item[Rule 4:]
\begin{equation}
\Delta G_i\,G^{-1}_i\,{\cal O}_i={\cal O}_i\,G^{-1}_i\,\Delta G_i
\rightarrow{\cal O}_i \, ,\label{rule4}
\end{equation}
\item[Rule 5:]
\begin{equation}
i{\cal Q}_i\,G^{-1}_i\,{\cal O}_i={\cal O}_i\,G^{-1}_i\,i{\cal
Q}_i \rightarrow 0 \, .\label{rule5}
\end{equation}
\end{description}
Hence  $\Delta G_i\,G^{-1}_i\to \bm{1}$ and $i{\cal Q}_i\,G^{-1}_i
\to 0$ for all operators ${\cal O}_i$ except ${\cal Z}^\ell$.

The one-body current operator satisfies
\begin{description}
\item[Rule 6:]
\begin{equation}
{\cal Q}_i J_i {\cal Q}_i\rightarrow 0\label{rule6}
\end{equation}
\end{description}

The algebra of these operators is nonassociative so it is
necessary to perform these operations by first applying rules 1--5
to eliminate all occurrences of the inverse one-body propagator
$G_i^{-1}$ then apply rule 6.

%==============REFERENCES=========================================

\end{document}